**The Concurrent Use of Medical Imaging Modalities and Innovative Treatments to Combat Retinitis Pigmentosa**

**Rickie Xian**
**University of Waterloo**
**r4xian@uwaterloo.ca**
**MSc Candidate**

## Abstract

Retinitis pigmentosa (RP), one of the leading causes of vision loss and blindness globally, is a progressive retinal disease involving the degradation of photoreceptors (7) and/or retinal pigment epithelial cells (14). Affecting approximately 1 in 4000 people, RP is caused by a series of genetic mutations, each specific mutation presents a specific pathological pattern in the patient, with the same mutation even presenting in different phenotypes in different patients (14). RP generally starts with peripheral vision loss, attacking the rods first, causing nyctalopia or night blindness (22). In later stages of the disease, the cones start to atrophy, further narrowing the field of vision and obscuring central vision (22). Luckily, with recent advances in medical imaging techniques and novel therapeutic treatments, both early detection and the overall prognosis of RP in patients have improved dramatically in the past few decades. This review will trace RP's physiological causes, how it affects retinal and ocular physiology, the techniques through which we can diagnose and image it, and the various treatments developed to try to combat it. The medical imaging techniques to be discussed include but are not limited to adaptive optics (AO), OCT including SD-OCT and OCTA, fundus autofluorescence (FAF) and its associated fluorescence lifetime imaging ophthalmoscopy (FLIO), colour Doppler flow imaging (CDFI), microperimetry, and MRI. The treatments to be discussed include stem cell therapy, gene therapy, cell transplantation, pharmacological therapy, and artificial retinal implants. Throughout this review, it will be made evident of not just the severity and diversity through which RP can present, but also the advances made in medical imaging and innovative treatments designed to combat this pathology.

## Introduction

Retinitis pigmentosa (RP) is a progressive retinal disease and one of the most common causes of acquired blindness worldwide, affecting roughly 1 in 4000 people (7). Caused by a series of genetic mutations that can occur through three Mendelian subsets: autosomal recessive, autosomal dominant, or X-linked, RP can present extremely different disease progressions depending on the specific pathological mutation as well as the patient themselves (14). The dominant forms of RP are characterized by more variable presentations with the best overall prognoses; recessive forms are characterized by intermediate presentation with modest prognoses, and X-linked forms are characterized by earlier disease onset and the worst prognoses (14). Regardless of the specific mutation harboured, all RP patients present a gradual photoreceptor loss (7). Clinically, in early stages of RP, the disease presents with peripheral vision loss and night blindness or nyctalopia due to rod photoreceptor cell damage, loss, and dysfunction (14). As the diseases progress, further changes to the retinal layer occur, often resulting in retinal pigment epithelial cell death (14): the patient will start to lose cone photoreceptor cells which progressively constricts the field of vision in concentric rings; at late stages of the disease, complete central vision loss occurs (14). Other symptoms include the reduction of the amplitude in electroretinograms (ERGs), bone spicule pigmentation (7), waxy optic disc pallor, and various morphological changes that occur in the retina pigment epithelium (RPE) (13). Though RP is a devastating disease with no known cure, recent advances in medical imaging and clinical treatments have greatly improved the overall prognosis of RP patients. Medical imaging techniques such as adaptive optics (AO), various forms of OCT including spectral domain OCT (SD-OCT) and OCT angiography or OCTA, fundus autofluorescence (FAF) and fluorescence lifetime imaging ophthalmoscopy (FLIO), colour Doppler flow imaging (CDFI), microperimetry, and MRI allow for the early detection and thus early intervention of RP. Clinical treatments have also been developed and shown to not just slow or stop disease progression, but also functionally recover vision (29), (32). In order to understand what causes RP, how we can diagnose and treat it, and what the prognoses look like in patients, this review will follow a sequential structure. First, background information will be provided on normal retinal physiology, the cause of RP, and the general setup and functionality of the imaging techniques presented for context. Then, the clinical symptoms and hallmarks of RP in the context of medical imaging will be discussed, elucidating how various techniques can be used for RP diagnosis and disease monitoring. Following imaging and diagnosis, the various treatments in development and their prognoses in patients will be presented, including stem cell therapy, gene therapy, cell transplantation, pharmacological treatment, and artificial retinal implants. Finally, the above information will be summarized, with specific emphasis on the importance and necessitation of the simultaneous and reciprocal development of medical imaging techniques and clinical treatments in order to effectively tackle this retinal disease which so many are afflicted with.

## Background

### Normal Retinal Physiology

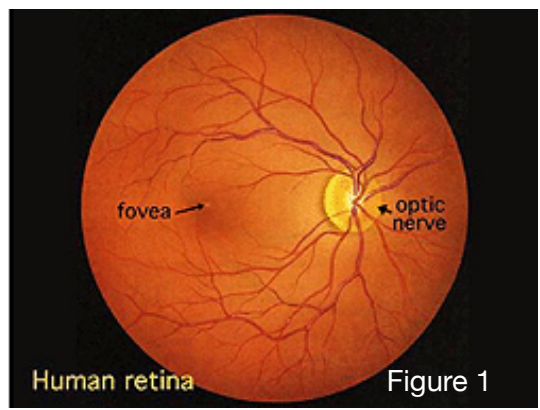

Figure 1

Before a meaningful discussion of RP's symptoms, diagnosis, and treatment, as well as how medical imaging techniques facilitate the above can be conducted, some background information must be provided for context. First, normal retinal physiology will be discussed for context of the morphological and functional changes caused by RP, as well as of the morphology-based treatments. Then, RP's genetic cause will be explained for context of gene-related treatments such as gene therapy. Finally, the general functionality and setup of the various medical imaging techniques to be used for diagnosis will be reviewed for context in the medical imaging and diagnosis section. Looking first at retinal physiology, the macroscopic view of the retina can be seen in Figure 1 (38). The ovular white area in the centre of the retina is the optic nerve, from which major retinal blood vessels radiate out (38). The slightly darker spot to the left of the optic nerve is the fovea, an area of the retina that sits in the central macular position where visual acuity is the highest (38). A more detailed diagram of the retina in the context of the entire eye can be seen in Figure 2. The fovea is surrounded by what's considered the central retina otherwise known as the central macula, a circular area of about 6mm; any area beyond this cutoff is known as the peripheral retina (38). A characteristic feature of the fovea is the foveal pit, an inward curvature at the centre of the fovea (38). The entire foveal area is known as the macula, which includes not just the foveal pit, but also the foveal slope, parafovea, and perifovea (38). The fovea is free of both innervation and blood supply as these structures would scatter the incoming light to the photoreceptors and thus greatly affect vision clarity (38). The absence of foveal interaction is achieved through ganglion cell axons from the ganglion cell layer lying above the inner limiting membrane (which will be discussed shortly) such that the central ganglion cell fibres project towards the optic nerve, circumventing the foveal slope; peripheral ganglion cell axons project along the same path (38). The absence of foveal blood supply is due to blood vessels ringing the fovea, but not existing within the fovea itself (Figure 1) (38).

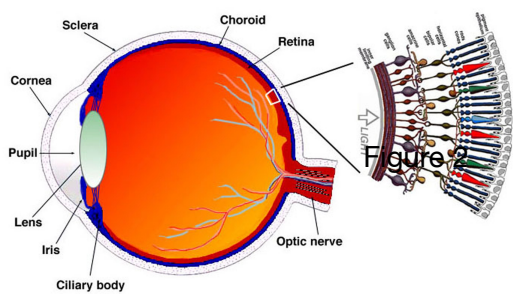

Figure 2

The retina itself, as seen in Figure 2, is congruent with the back of the eye and relatively thin at about 0.5mm in width; it's partitioned into an organized series of layers, each with a specific function (38). Ganglion cell axons which connect to the brain along with blood vessels for retinal and neuronal vascularization are both contained in the optic nerve; from Figure 2, it's clear that the neurons responsible for retinal neuronal output, the ganglion cells, exists in the retinal layer closest to the lens, whereas the photoreceptors cells exist in the retinal layer furthest from the lens, adjacent to the choroid and retina pigment epithelium (RPE) (33), (38). As a consequence of this physiological organization, light coming in from the pupil must traverse the entire retina before it can come into contact with and activate the photoreceptors (38). There are two types of photoreceptors within the eye, rods, cells activated in low light level vision (scotopic vision) but not capable of colour vision, and cones, cells activated in high light levels (photopic vision) for producing colour vision and spatial acuity (14). Rod cells comprise approximately 95% of human photoreceptors, with cones comprising the remaining 5% (7). Once the photoreceptors absorb the photons through visual pigments, the vision cascade is triggered, converting biochemical stimuli into an electrical impulse for the stimulation of the ganglion cells (38). The ganglion cells then emit a pulsating neuronal pattern, informing on basic light information and image organization, which then gets relayed to the brain (38). A more detailed view of the retinal layers and organization can be seen in Figure 3, where the complex organization and perfusion of nerve cells in the retina can be appreciated; the retina itself actually contains three nerve cell layers, the outer layer housing the photoreceptors, the middle layer hosing the bipolar, horizontal, and amacrine cells, and the inner layer housing ganglion cells and any displaced amacrine cells (38). The individual function of each neuronal layer (excluding the photoreceptor layer) and how they relate with one another however is not greatly significant to this review, so it won't be discussed in further detail.

Figure 3

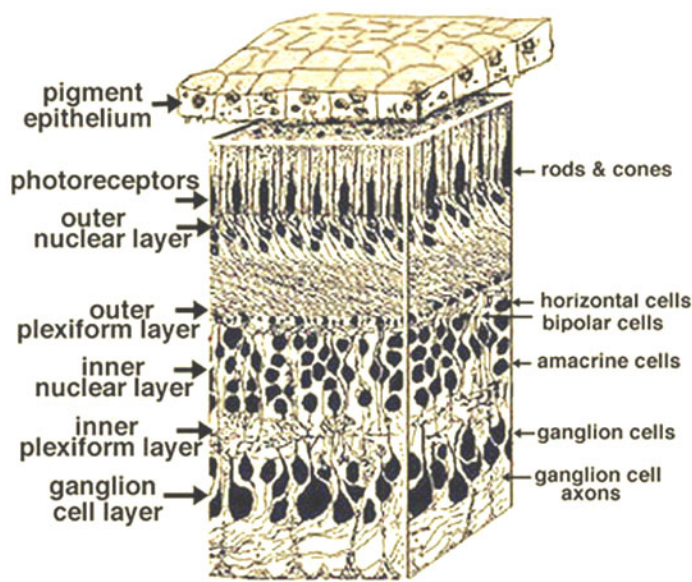

Though understanding each retinal layer's general function is important, it is of equal if not greater importance to understand how each retinal layer is defined and what it's composed of. Thus, each retinal cell layer's composition will be explicitly explained for context of the results presented in the Imaging and Diagnosis section of this review. The pigment epithelium, otherwise known as the retinal pigment epithelium or RPE, is the pigmented cell layer on the outermost retina (Figure 3), which is connected to Bruch's membrane (the innermost choroid layer) and the choroid itself (38). The retinal pigment epithelium topography is achieved within the optic nerve, which connects the visual cortex and the lateral geniculate (38). The choroid itself is responsible for 65-85% of the blood flow, nourishing the outer retinal layer, specifically the photoreceptors; the remaining blood flow is directed through the central retinal artery, supplying blood to the optic nerve head and inner retinal layers (38). The retinal arterial blood flow is known to be locally, metabolically regulated, whereas the choroidal blood lacks this regulation (57). Four main branches comprise the central retinal artery; these branches then supply three capillary network layers (38). Plunging deeper into the retina, it can be seen the next layer is the outer limiting membrane or OLM, composed of adherens junctions between photoreceptor inner cell segments and Muller cells (radial glial retinal cells), forming a barrier between the neural retina proper and the subretinal space (38). The layer following the OLM is the outer nuclear layer (ONL) composed of rod and cone cell bodies; the next layer is the outer plexiform layer (OPL), which houses the synaptic connections between the photoreceptors and the bipolar and horizontal cells (38) (Figure 3). The inner nuclear layer (INL) exists under the outer OPL, containing the horizontal cells, bipolar cells, and amacrine cells; the bipolar cells then synapse onto the ganglion cells in the inner plexiform layer via the ganglion cell axons (the IPL) (Figure 3) (38). The next layer of the retina is the ganglion cell layer (GCL) where, as the name suggests, the ganglion cells are located; the final, innermost retinal layer is suitably defined as the inner limiting membrane or ILM, a layer which houses Müller cell end feet which are laterally connected, lining the vitreous humour as the retina's inner surface and acting as a natural barrier to diffusion between the vitreous humour and neural retina (38).

Now that an overview of retinal physiology, its layers, and its functions has been conducted, appreciable differences between the central and peripheral retina will be discussed, as these distinctions are highly relevant to this review. First, the central and peripheral retina differ in their photoreceptor cell type composition: the central retina contains much more cones than rods, in contrast to the peripheral retina, which houses the inverse situation of much more rods than cones (38). Due to the higher density with which photoreceptors (notably cones) and their affiliated ganglion and bipolar cells pack within the central retina versus the peripheral retina, the central retina is much thicker; this higher density also accounts for the thicker inner nuclear layer (INL), inner plexiform layer (IPL), and ganglion cell layer (GCL) in the central retina versus the peripheral retina (38). Interestingly, there exists no remarkable difference in thickness between the outer nuclear layer (ONL) in the central and peripheral retina as the ONL is composed of rod and cone cell bodies, however the composition of photoreceptor cell types differs greatly between the two regions as previously mentioned (38). A related feature to distinguish the central retina from the peripheral retina is that cones within the central retina are displaced from their synaptic stems due to the oblique axons associated with the cones (38). The Henle fibre layer, a fibrous region that stains subtlety with exogenous dyes, is in fact formed by these oblique axons, a feature the peripheral retina lacks (38). Now that an overview of retinal physiology has been conducted, the cause of RP can be discussed with more context.

**Retinitis Pigmentosa Causes**

It's known that RP is caused by a series of genetic mutations, however its not known exactly how each of these mutations actually trigger retinal cell death. First, the known number of mutations that can cause RP are within the hundreds (11), as there exist over 150 genes associated with RP (14). Second, the same gene mutation can manifest in a variety of ways in different patients depending on their diet, environment, lifestyle, etc (11). Finally, notably, there exist two subcategories of RP, syndromic and asymptomatic, in which the former affects other tissues and organs besides those of the retina, and the latter is localized to only the retina (14). These RP-affiliated gene mutations can be roughly categorized into three Mendelian subtypes as mentioned in the introduction: X-linked, autosomal recessive, and autosomal dominant; these three subtypes are ordered in descending order of disease onset intensity and ascending order of overall prognosis, with X-linked forms being the most intense with the worst prognoses, and autosomal dominant forms being the least intense with the best prognoses (14). The types of genes susceptible to RP-associated mutations are multitudinous, with some common types being: the eyes shut homolog (EYS) gene, mutations of which are associated most commonly with autosomal recessive RP (11); the retinal pigment epithelium (RPE) gene, mutations of which are most commonly associated with autosomal dominant RP (12), the rhodopsin, peripherin, and cGMP phosphodiesterase genes (17), and the retinitis pigmentosa GTPase regulator (RPGR) gene and RP2 gene, mutations of which are most commonly associated with X-linked RP (24), (26). As mentioned, though the exact mechanisms by which these mutations initiate cell death is vaguely understood at best, meaningful attempts have been made to deduce them. Exceptionally, the work done by Portera-Cailliau et. al involving agarose gel electrophoresis and in situ apoptosis cell labelling showed in three mouse models of RP caused by peripherin, cGMP phosphodiesterase, and rhodopsin mutations, that photoreceptor cell death occurred through traditionally apoptotic mechanisms in place of necrotic mechanisms (17).

An interesting point to be made is that intrinsically disordered proteins or IDPs are often involved in the disease cascade (25). IDPs themselves are proteins that lack a fixed tertiary structure; instead of the conformational changes triggered by certain stimuli, for example changes in temperature, pH, or ligand binding, in traditional proteins, IDPs seem to spontaneously sample a wide array of conformations (2), (3), (4). Traditional proteins usually contain a few well-defined local free energy minima corresponding to stable structures, whereas IDPs have a much more jagged free energy landscape with many local minima; thus IDPs can be thought to diffuse through their energy landscapes, allowing them to easily sample their many different configurations (2), (3), (4). IDPs are involved in essential cellular functions like signalling or translating, and are found to be relatively abundant within the eye's aqueous humour (1), (6), as well as within ribonucleoproteins (27). A salient experimental example of IDP involvement in RP disease mechanisms was conducted by Picarazzi et.al, who found that the gene P347 of rhodopsin mutation linked RP contained an intrinsically disordered region within the C-terminus (25). By using MD (molecular dynamics) simulations, Picarazzi's group was able to determine the most probable cause of pathogenicity to be the intrinsically disordered region within the mutant rather than other steric or physiochemical determinants (25). With the brief summary of the genetic cause of RP concluded, now comes the section through which the setup and methodology of the imaging techniques used to image and diagnose RP will be discussed.

**Medical Imaging Techniques for Retinitis Pigmentosa Imaging**

As a slight tangent before the formal review of medical imaging can be conducted, electroretinograms (ERGs) are a medical technique worth mentioning, as they are often used supplementarily with imaging techniques for retinal imaging (46). Conventional ERG systems are composed of a light source, electrodes, an amplifier, a computer, and a camera (56). In essence, the ERG measures the retinal global electrical response due to light simulation, using it to assess retinal pathology and function (56). The eye is illuminated with some bright light source, resulting in voltage versus time data which can be recorded by electrodes placed at the cornea (46), signal amplified, then displayed (56). The electrical waveform traditionally contains two features: the a-wave, the large negative electrical component or "dip" in the voltage-time graph, and the b-wave, the higher amplitude, positive electrical impulse (46). The a-wave is known to inform on outer retinal photoreceptor health, whereas the b-wave informs on inner retinal health, for example the health of Müller and ON bipolar cells (46). Though sometimes two other waveforms, the c-wave and d-wave, are recorded, further detailed discussion of ERGs is unnecessary for this review. A newer implementation of ERGs is multimodal ERGs (mfERG), useful for diagnosing unknown vision loss, differentiating between the similarly clinically presenting retinal and optic nerve diseases, and determining Plaquenil (hydroxychloroquine) toxicity (56). In lieu of the simple light source in ERGs, mfERGs use a complex monitor, which stimulates the patient's retina with a pattern which changes quickly and quasi-randomly; cross-correlation software then deduces segmented retinal response, allowing for a more detailed understanding of the retina's functionality (56).

Now, medical imaging techniques can be traced. The list of techniques to be presented in this review is not exhaustive, and includes adaptive optics (AO), various forms of OCT including spectral domain OCT (SD-OCT) and OCT angiography or OCTA, fundus autofluorescence and fluorescence lifetime imagine (FLIO), Doppler flow imaging (CDFI), microperimetry, and MRI. First, adaptive optics (AO) imaging will be discussed, a technique which is used in tandem with other imaging systems; the most characteristic feature AO employs is the usage of a mirror capable of undergoing controlled deformations within an imaging system to correct for aberrations and improve spatial resolution (34). The eye naturally contains many optical imperfections, such as those arising from the structural diversity of the tear film, cornea, and lens (34). These variable optics are deemed optical wavefront aberrations, and cause light beams to strike the retina at different locations depending on the location through which they entered from the pupil; these monochromatic wavefront aberrations will result in a noticeable blur or distortion in distant light sources (34). Chromatic aberrations also affect the spatial resolution of retinal imaging systems, as these arise from the varying refractive index the eye has depending on the incoming wavelength of the light beam (34). By employing AO in imaging systems with near-infrared wavelengths, researchers can safely and non-invasively correct for the distortions in the light beam entering the eye by carefully altering the mirror's surface, allowing for light beams which enter at different pupil positions to strike the retina in the same location (34). The implementation of a wave sensor, for example a Shack Hartmann (SH) wavefront sensor, then further improves the aberration correction, as it allows for detailed wavefront information to be sensed and recorded such that the mirror shape can be altered precisely for maximal resolution (34). The manner in which the SH sensor functions is noteworthy and should be discussed in brief; an array of tiny lenses or lenslets comprise these sensors, which work to focus collimated light onto an area sensor, for example a CCD or CMOS camera (34). In retinal imaging, as light rays depend on pupil location, a point source illumination of the retina produces a bright spot, which then produces a wavefront with a curved surface; when the wavefront reaches the SH sensor, each lenslet receives a wavefront with a different angular position (34). It's known that the image produced by a lens is tilted at an angle relative to the optical axis which is proportional to the incoming beam angle, producing bright spots on the sensor wherein the displacement of the spot from each lenslet's optical axis is proportional to each lenslet's detected wavefront slope (Figure 4).

**colour  Figure 4**

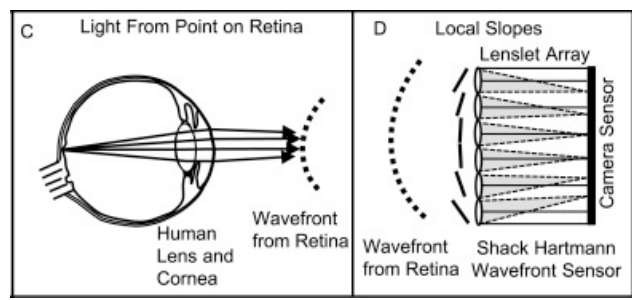

AO can generally be partitioned into three categories: sensor based hardware AO (HAO) which was discussed above, sensorless AO (SAO), and computational AO (CAO) (34). The schematics of each type of AO system can be seen in Figure 5: HAO uses hardware for both wavefront sensing, with a wavefront sensor, and wavefront shaping, with a deformable mirror; SAO lacks a wavefront sensor, and CAO lacks both a wavefront sensor and a wavefront shaper (34). Though HAO is the most common type of AO used in retinal imaging for its ease in software operation, the hardware is costly and complex; to overcome these disadvantages, SAO and CAO can be used (34). In place of a physical wavefront sensor, SAO uses the image properties themselves to estimate wavefront quality and inform the wavefront shaping device (34). An image metric is calculated each time a deformation is made to the mirror, allowing for iteration over the metric-deformation loop until a diffraction limited image is converged upon; this iterative technique allows for the production of the best aberration-corrected image, as image quality and aberrations are inescapably linked (34). In contrast, CAO circumvents both wavefront sensing and wavefront shaping hardware with the implementation of a measurement sensitive to phase and amplitude of the retuning light; these data are then processed with a computational algorithm which allows for efficient convergence to the necessary diffraction-limited alteration for wavefront correction (34). The system then converts the calculated correction into a digital filter which the input image then passes through, allowing for aberration corrections without the need for wavefront hardware (34). Regardless of the type of AO implementation, all AO systems involve a source illumination, waveform sensor and wavefront shaper (be it hardware or software), and a fundus camera for image capture (34), (52).

Figure 5

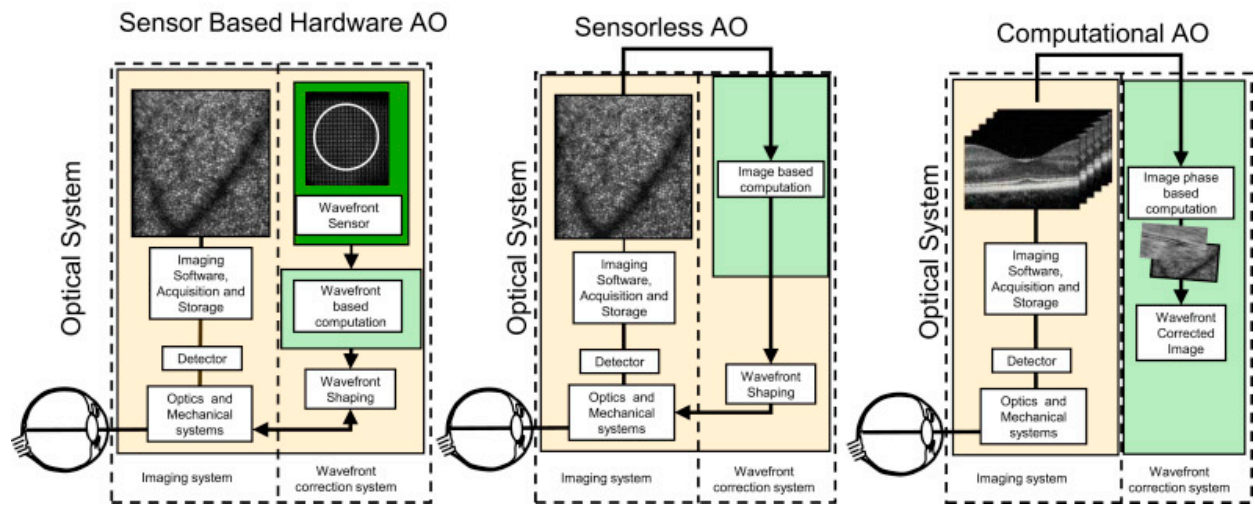

The illumination method of AO is also noteworthy, as AO uses either scanning confocal imaging systems such as scanning laser ophthalmoscopy (SLO) systems or flood-illuminated systems (52), and can often be used to enhance other imaging modalities such as OCT, deemed AO-OCT (34). Before discussing OCT functionality, the distinctions between scanning confocal imaging AO and flood-illuminated AO should be made. In confocal imaging systems, the retina is illuminated with a low illumination power, (52) focused light source which scans or sweeps across its entirety in a raster pattern (55); the sample-reflected light passes through a beam splitter, then gets focused onto an optically conjugate confocal aperture relative to the imaging beam's focus plane (34). The confocal aperture thus re-focuses and re-images the imaging beam's back-scattered light before detection, such that each signal denotes a specific location in the retina (34). Back-scattered light from structures outside of the focus plane are excluded from the image, meaning confocal AO systems have higher image contrast as out of focus light has no effect on the image (34), allowing the system to "axially section" the image (52). In contrast, in flood illuminated systems, the entire target retina area is illuminated in one continuous beam, with the source light beam entering through the pupil (34). Back-reflected light from the sample is imaged onto the camera which is once again, optically conjugate to the focus plane of the imaging beam (34). Thus scattered light from other areas of the eye are projected onto the camera, but remain out of focus; this type of AO illumination is known to be diffraction limited, wherein the angle of the pupil's subtended light relative to the retina dictates the image resolution (34). With the comprehensive review of AO completed, and the mention of AO-OCT as an imaging modality AO is often added to to enhance imaging, the paper now allows for a smooth transition into the explanation of OCT imaging systems.

OCT or optical coherence tomography in a nutshell is an imaging technique analogous to ultrasound, but instead of using sound waves, it uses light waves (22). It is an extremely well established ophthalmological imaging technique often used for in situ and in vivo retinal imaging (22), capable of producing high resolution, cross-sectional tissue structure images (40). OCT was initially introduced as time-domain OCT (TD-OCT), and should be carefully distinguished from spectral-domain OCT (SD-OCT), an explanation of which will shortly follow. Returning to the analogy between ultrasound and OCT, both imaging techniques subject a tissue to waves; the back-reflected waves "echoing" off the tissue can be analyzed wherein the "echo delay" can be used to trace reflection depth and thus reveal structural information of the tissue examined (40). OCT often uses light in the near-infrared range for resolution and safety purposes; because the speed of light exceeds the speed of sound by many orders of magnitude, back-reflected light wave delays can't be measured directly (40). As an alternative, if one were to have a reference length, an indirect measurement of the delay can be made (40). OCT accomplishes this task by using an interferometer, which splits the light source beam, sending one beam to the tissue sample, and another to a known length reference arm (Figure 6); in fact, low-coherence interferometry is the central tenet to all forms of OCT (40). Any light source has a certain temporal coherence, a property which defines how continuous a wave train is at any position; low-coherence light sources have a wide bandwidth and a short coherence length, defined as the length over which the light remains coherent with a constant phase relation (40). Circling back to the reference beam and sample beam in OCT systems, the interferometer then recombines the two beams, wherein the interference of the two and the associated light intensity are digitally detected with a photo detector (Figure 6) (40). Low-coherence light only interferes when the optical path lengths of both the reference and sample beam are equal and in the coherence length range; in this way, the reference path length can be altered to specific depths such that only those sample beams which match the reference length interfere (40). Consequently, imaging depth can be selected for by for example using a reference mirror to change the reference length and record successive depth profiles or A-scans (Figure 6) (40). The above implementation is the oldest and most standard of OCT implementations, and is deemed time-domain OCT (TD-OCT); in order to generate cross-sectional images or B-scans, the sample must be illuminated laterally by the sample beam (40).

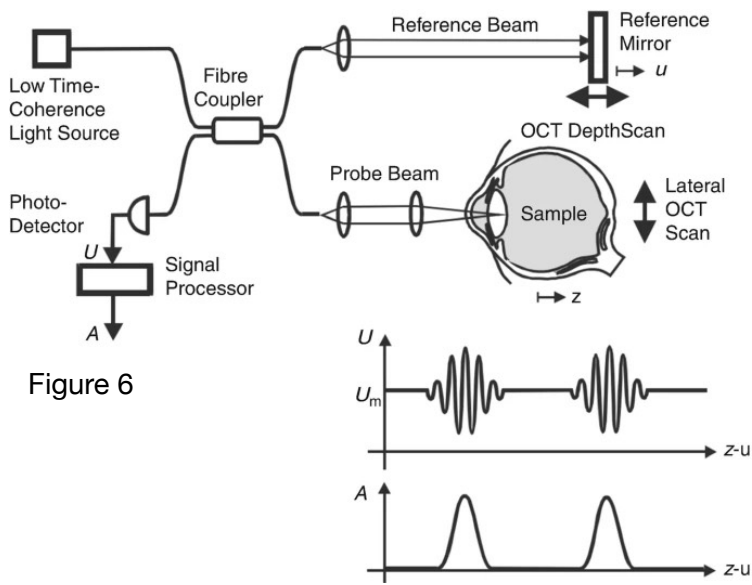

Figure 6

Figure 7

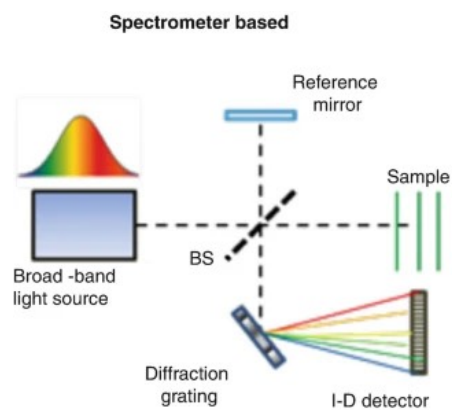

After TD-OCT was introduced, the development of Fourier domain OCT (FD-OCT) followed shortly thereafter, a technique which more efficiently exploits low-coherence interferometry; in lieu of optical path length mechanical scanning, FD-OCT captures A scans solely with spectral data (40). Spectral domain OCT (SD-OCT), otherwise known as spectrometer based FD-OCT, uses a setup similar to that of TD-OCT, except a spectrometer stands in place of a point detector, a diagram of which is illustrated in Figure 7 (40). A line image can consequently be split into its constituent wavelengths, a task accomplished by a diffractive element housed within the spectrometer; a high speed line scan camera records the wavelength-separated data, wherein each camera readout corresponds to a spectral interferogram with specific fringe patterns (40). Notably, interferograms inform simultaneously on all known depth information of a given sample; thus, the interference signal must be Fourier transformed to sequester each depth layer's contribution as a function of intensity versus depth (Figure 8) (40). As an aside, a mirror term is produced within this transform due to the Fourier transform's function's symmetry, which is excluded from the final image (40).

Figure 8

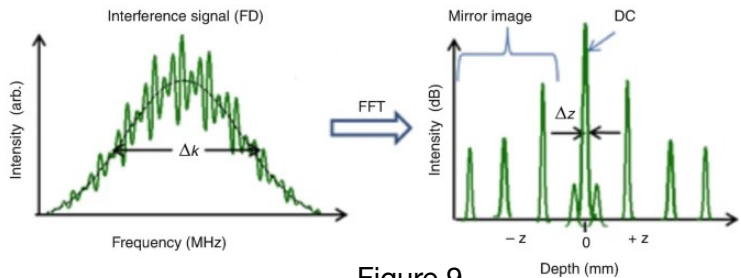

Figure 9

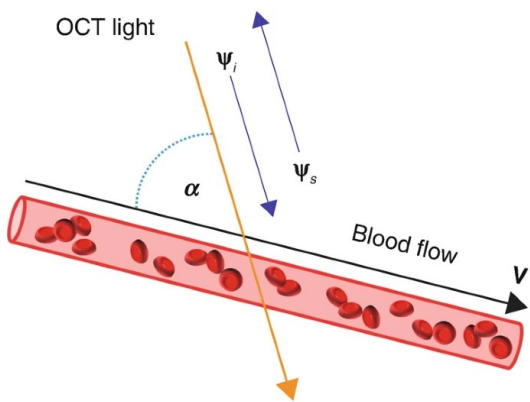

Wrapping up this subsection on OCT will be the modest explanation of OCTA, OCT angiography. OCTA, a non-invasive alternative to conventional fluorescence angiography (9), is used for visualizing vasculature, most commonly the posterior eye blood vessel network (42); this derivative of OCT uses the difference in the intensity or amplitude between consecutive, cross-sectionally identical OCT B-scans to construct a 3D blood flow "map" (41). To ensure motion between B-scans is solely due to red blood cell motion, the patient is spatially restricted such that axial bulk motion is nonexistent (41). A more detailed schematic and explanation of how OCTA uses blood flow velocity measurements to image 3D vascularity is seen in Figure 9: an incident OCT beam strikes the blood vessel, and is then quickly scattered by the blood cell's motion (42). The angle $\alpha$ is the angle between the blood flow velocity V direction and the incident OCT beam; the moving blood cells partially reflect the incident light wave $\psi_i$, directing the reflection into the scattered light wave $\psi_s$ ((Figure 9) (42). By inspecting the scattered light wave modulation from consecutive B-scans, one can calculate the blood cell velocity (42). Thus, OCTA informs on flow information at some point in time, wherein this flow information intrinsically contains both functional and structure information (41). OCTA is sometimes supplemented by additional quantitative OCT measurements for more direct blood flow velocity determination. Despite this supplementation with conventional OCT, OCTA serves as an effective tool for blood flow detection such that retinal vasculature can be clearly visualized, especially the anastomosing blood and lymphatic capillary network between the IPL and INL, the superficial capillary plexus (SCP), and the deep capillary plexus (DCP) (9).

The next imaging technique to be discussed is fundus autofluorescence (FAF) and its associated fluorescence lifetime imaging ophthalmoscopy (FLIO). FAF is a non-invasive, retinal imaging modality capable of detecting certain fluorescent compounds which occur naturally in the retina; these compounds or fluorophores become photo-excited, allowing for their electrons to jump to an elevated energy state (43). The electrons then transition back to ground state, releasing energy in the form of wave-lengthened light of a lower energy than the excitation wave; the predominant fluorophore located within the macula which FAF functionality relies on is lipofuscin, a molecule that emits yellow-green light in the 600-610nm range (43).

Due to lipofuscin's peak excitation at 470 nm within the blue-green spectrum, researchers often use blue-green light as their excitation bandwidth, otherwise known as short-wave fundus autofluorescence (15); it follows that wavelength can be modulated to detect other fluorophores with different peak excitations (43). Bisretinoids, metabolic byproducts of the visual and vitamin A cycle, are responsible for the autofluorescence seen in lipofuscin; these byproducts form initially within the photoreceptor outer segment before being converted to lipofuscin for RPE deposition (43). Lipofuscin RPE lysosomes concentration is directly proportional with age and severity of degenerative disorders such as age-related macular desecration (AMD), macular dystrophies such as Stargardt disease (43), and of course, RP (15), (18). Thus, FAF captures the spatial intensity of the fluorescent emissions, producing the necessary information to form a map of the lipofuscin distribution (43).

Often times in FAF, not just the intensity of the fluorescent emissions is of interest, but also the time the fluorophore spends in its excited state, otherwise known as the fluorescent lifetime; these lifetimes vary with metabolic states, thus indicating retinal molecular state and disease progression (18). To measure and record these fluorescent lifetimes, fluorescent lifetime imaging ophthalmoscopy (FLIO) is often used in tandem with FAF, a method which exploits time-correlated single photon counting (TCSPC) (18). TCSPC's central principle is that in the brief delay after excitation, only one fluorescent photon is detected; as such, the probability of multiple photon simultaneous detection is low (44). Thus, by detecting photons as a function as time, researches can construct the proportional probability density of photon emission, otherwise known as the dynamic fluorescence, and derive the fluorescent intensity as a function of time (44). A schematic of a cLSO (confocal laser scanning ophthalmoscope) system which employs FAF and FLIO can be seen in Figure 10, providing a brief overview of the experimental setup (44).

Equation 1

$$f_{\mathrm{d}} = \frac{2f_{\mathrm{t}}|v|\cos\Theta}{c}$$

With the discussion of FAF and FLIO concluded, colour Doppler flow imaging (CDFI) can now be presented. CDFI is a subset of ultrasound imaging commonly used for blood flow visualization; in conventional ultrasound imaging, a pulse-echo (PE) scheme is executed to image tissue cross sections however with CDFI, an additional colour map is overlaid onto the PE image, where the colours denote movement (45). Generally, CDFI relies on physical principles similar to those of PE ultrasound imaging: ultrasound beam pulse transit time and direction data are used to determine the structural location of the imaged tissue to construct each image's pixel, but instead of amplitude analysis of the echoes, Doppler shift analysis is conducted (45). A subtle point to be made regarding this Doppler analysis is that the Doppler shift isn't calculated for each pulse; instead, the phase shift time delay between concurrent echoes of the same volumetric location is estimated (45). The phase rate of change can then be denoted as a frequency shift such that the conventional Doppler shift formula (Equation 1) can be used to calculate the target velocity (45). Looking more closely at Equation 1, $f_{\mathrm{d}}$ represents the Doppler shift frequency, θ is the angle between the flow vector and the ultrasound beam, $|v|\cos\Theta$ is the vector component of the velocity in the direction of the transducer, $f_{\mathrm{t}}$ is the transmitted ultrasound frequency, and c is the ultrasound velocity in tissue (45). In conventional CDFI implementations, the target velocity in the direction of the transducer is directly measured, meaning detailed directional flow information must be known for quantification of the velocity vector (45). It should be noted that though CDFI and PE ultrasound imaging overlap in their principles, CDFI is more technical in its implementation; the explanation behind this technical difference is twofold: blood flow targeted by CDFI contains echoes with magnitudes significantly less than those of solid tissue (the conventional PE ultrasound imaging target), and CDFI targets must undergo multiple pulses in order to properly estimate the velocity, compared to PE ultrasound requiring only a single pulse (45).

The conventional CDFI imaging system setup is array transducer based, and illustrated in Figure 10, wherein RF denotes radio frequency, TGC time gain compensation, I the signal in-phase components and Q the signal quadrature components (45). Going though the essential element's functionality in Figure 10: a multitude of elements, each capable of ultrasound pulse transmission and reception, comprise the transducer; the beam former serves a dual purpose, correcting the sequence and combination of transducer elements such that an appropriate transmitted beam can be formed, and combining the retuning echos for approximate receiving beam generation; the RF amplifier serves to amplify the beam former output while undergoing time gain compensation such that high depth echos' excessive attenuation can be compensated for; the quadrature demodulator then demodulates the Doppler signal into in-phase (I) and quadrature (Q) components which are then stored in colour vector memory; the signal is then filtered and processed such that each volumetric Doppler signal echo can be quantified into its power, mean frequency, and bandwidth; then, final processing occurs, allowing for PE imaging data to combine with the Doppler signal data, resulting in a simultaneous visualization of velocity and anatomy (45). Laser Doppler flowmetry, a modification of CDFI wherein a laser is used as the light source and the laser Doppler shift is measured, is a subset of CDFI and is also often used in the implementation of CDFI (59).

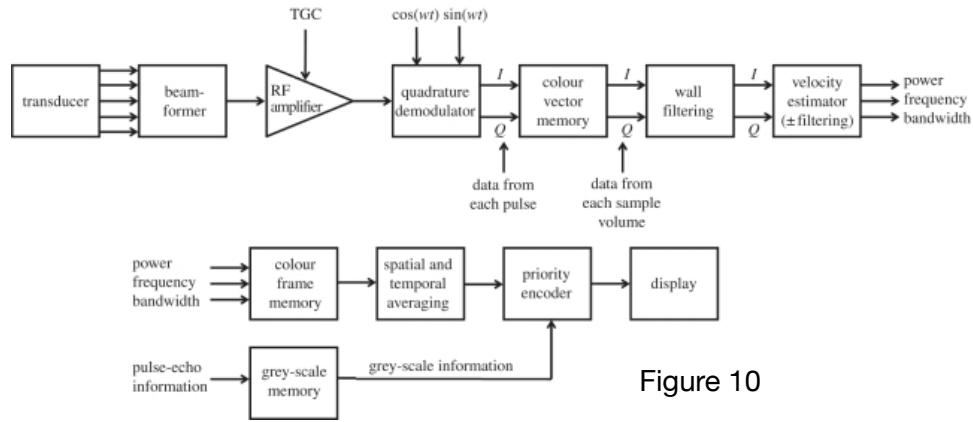

Figure 10

After the above brief summary of CDFI, the second to last imaging technique to be covered is microperimetry (MP), sometimes referred to as fundus-guided perimetry (37). MP is used to visualize and assess central retinal function using static-automatic perimetry (36), wherein the patient focus on a fixed target, and retinal sensitivity is measured by tracking eye movement (22). The conventional setup and sequential functionality of an MP system can be seen in Figure 11 (46). By illuminating the tissue and using a fixation target (46), researchers can use retinal sensitivity data and eye movement or fundus tracking to correctly execute the functionality-physiology correlation (47). The technique allow for exact correlation of any retinal location's sensitivity threshold with its clinical appearance in real time; the fixation, or maintenance of an eye's focus on a stationary location, is also recorded parametrically as location and stability data (47). Though in previous decades, MP relied on scanning laser ophthalmoscopes (SLOs) for movement tracking (47), (48), current MP systems commonly use fully automated microperimetres such as MP1 microperimetres, housing software capable of autonomous eye movement tracking (47). A reference frame of baseline ocular position is used to record the eye's shifts in the x-y coordinate system; then, the current frame's fundus position can be used to correct the location of the stimulus in real time (47). Fixation is recorded as a continuous parameter by default during a single MP exam, but can also be recorded in isolation, representing static fixation; an embedded CCD camera then captures a colour fundus photo, which the generated fixation data is then precisely overlaid onto (47). Each retinal test point's retinal sensitivity is recorded and represented as pseudocolours, which can then be used to construct a sensitivity map to be overlaid with a retinal image (for example one captured with concurrent OCT) (47). The ability of MP to stimulate the retina at precise locations while continually visualizing it during and between examinations is the backbone of its utility and efficacy in fundus tracking (46). The intrinsic method in which MP continuously collects data and tracks eye-movements allows for retinal sensitivity testing in the absence of stable fixation, repeated sensitivity testing of the same retinal location for increased reliability and re-testability, and correlation of structure and function (46). Figure 11 effectively summarizes the instrumentation and methodology executed by MP, as well as provides the background luminance of scotopic, mesopic, and photopic vision for reference (46). The imaging unit is generally an infrared SLO or fundus camera, which then feeds into the control system where the above mentioned fundus-tracking and stimulus positioning correction occurs; the control system then feeds into the stimulus and background unit, which is generally LCDs or LEDs;

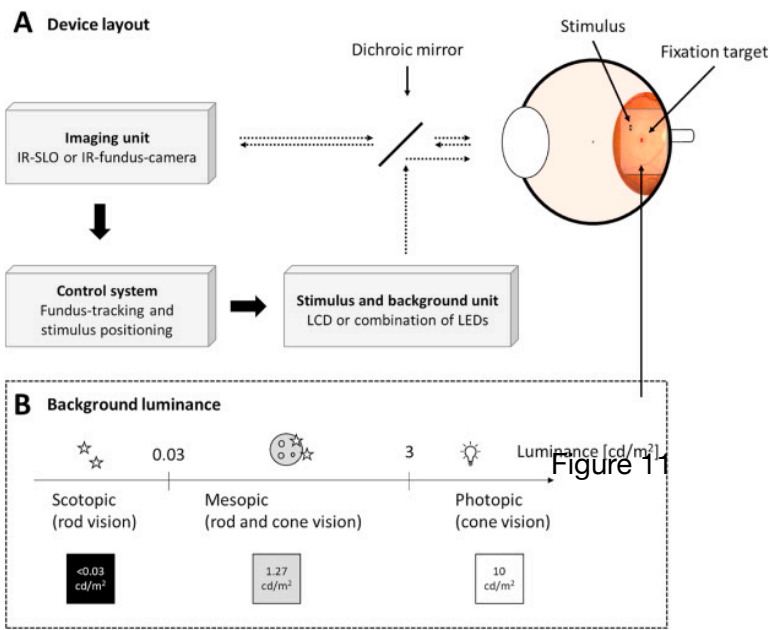

Figure 11

**Equation 2**

$\Delta L / L = \text{constant}$

these illumination arrays then reflect off a dichroic mirror for retinal stimulation for data collection (46). As a theoretical aside, the decibel scale and its implementation in MP is worth discussing in short. The decibel scale is predicated off of Weber's law, which states that the smallest change in stimulus that can be perceived ΔL, otherwise known as the just-noticeable difference, and the initial stimulus intensity L are directly proportional according to Equation 2 (46). With the implementation of the logarithmic decibel unit wherein the threshold differential luminance ΔL is defined as the difference in threshold and background luminance relative to the maximum stimulus luminance L_max capable by the illumination instrument, MP adopts a rather counterintuitive

**Equation 3**

$$\Delta S \ (in \ dB) = 10 \times \log_{10} \frac{L_{max}}{\Delta L}$$

brightness scale; as a mathematical consequence of Equation 3 (and indeed of any logarithmic measurement system), a reading of 0 dB corresponds to the maximum device luminance (46). Each subsequent 1 dB increase therefore attenuates the stimulus luminance by an approximate factor of 1.26 (46).

**Equation 4**

$\omega = \gamma B\_o$

The final imaging technique for RP to be outlined is perhaps the most colloquially known imaging technique, magnetic resonance imaging or MRI. Though MRI isn't generally used for ocular imaging, the next section will demonstrate a few innovative implementations of MRI researchers have used for RP imaging (20), (35). MRI functionality relies heavily on an intrinsic property of specific atomic nuclei: spin; hydrogen nuclei, bound to tissue and unbound as "free water", are generally the most relevant in the context of biological tissue, which is water-heavy (49). Nuclear spin is the mathematical analog to a nucleus with a positive charge due to its protons spinning on its axis (49). Note no mechanical spinning of the nuclei actually occurs, rather a magnetic field and dipolar magnetic moment are generated and induced simply due to the intrinsic nature of atomic nuclei (49). When an external magnetic field B_o is applied as in MRI, magnetization of the magnetic moments M_o of the hydrogen nuclei occurs (50), which causes most of the magnetic moments to align in the direction of the B_o field (49). The angular momentum of the nuclei due to their rotation causes them to precess about the $B_o$ axis (Figure 12), wherein the rotational velocity of the precession is known to be the Larmor frequency, which is directly proportional to the external B-field, related by the gyromagnetic ratio γ as seen in Equation 4 (49). The MRI system then excites the nuclei with the application of an additional radiofrequency B1 (RF) perpendicular with respect to $B_o$ (49); these RF are pulsed such that their frequency matches the Larmor frequency ω (otherwise known as the resonance frequency) (50). As the nuclei absorb the RF energy, they jump up to a higher energy level (49), interfering with their previous $B_o$ alignment precession by tilting their axes of rotation (50).

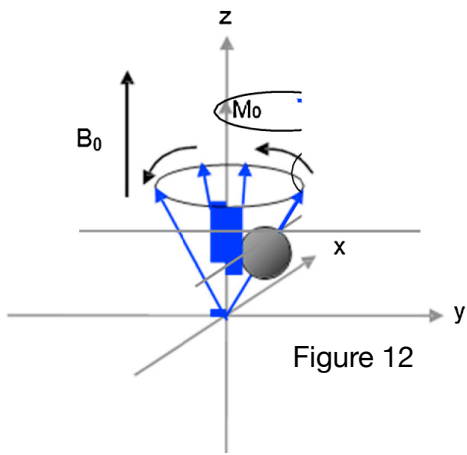

Figure 12

Figure 13

Eventually, the precessing nuclei transition back to their ground state in a characteristic biphasic relaxation pattern, emitting energy (50). The nuclear energy absorption and emission is known to be the MRI signal or "free-induction decay FID" (49), with the measure of coherence in the precession phase being directly proportional to signal strength (50). The two-fold process of the relaxation is partitioned such that the precession coherence loss with respect to time is denoted the T2 relaxation, and the nuclei orientation returning to their ground state or thermal equilibrium of $B_o$ field alignment is denoted the T1 relaxation (50). The energy change transfers within the system are detected by a wire coil using Faraday induction of the electromotive force (50), then amplified into the FID (49). For the improvement of the signal to noise ratio (SNR), MRI systems excite the tissue with numerous RF pulses, then average their corresponding FIDs (49). Three electromagnetic gradient coils in all cartesian directions (x, y, and z) are used to alter the applied magnetic field producing a magnetic gradient in any direction which is both phase encoded and frequency encoded (49) (50). As the precession rate is directly proportional to MRI signal magnitude, researchers can correlate the signal strength to differential spatial locations (49). Moreover, volumetric slices can also be selected for with the implementation of slice-relative orthogonal gradients, which alter the various nuclear precession rates such that only the target volumetric slice has a resonance frequency matched by the RF pulse (50). Thus the application of a known magnetic gradient allows researchers to reconstruct 3D images (49) and target specific volumetric segments within the imaged tissue (50). The final step for signal processing before image display is the Fourier transform of the averaged FID from the time domain into the Fourier domain; alternatively, the Fourier transform can be used to glean the frequency spectrum, which informs on the biochemical nature of the imaged tissue. The setup and process of image formation discussed above is summarized nicely within Figure 13 (51). Thus concludes this subsection on background of normal retinal physiology, retinitis pigmentosa causes, and the various medical imaging techniques used to image and diagnose RP. At this point in the review, the necessary context for incisive interpretation of the imaging and diagnosis and treatment subsections to follow has been provided.

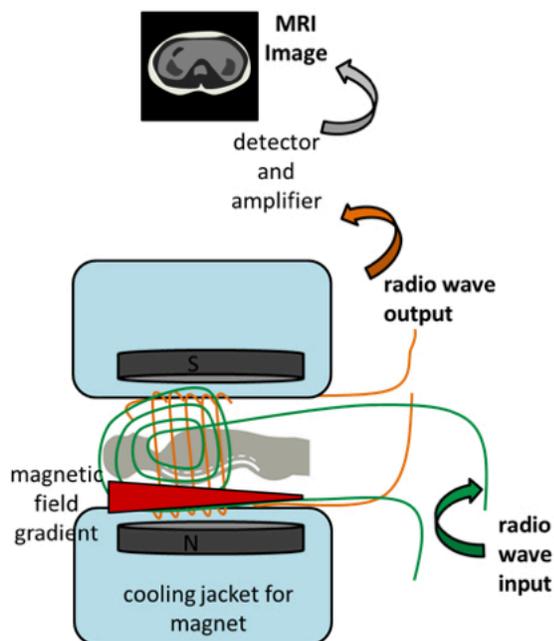

**Imaging and Diagnosis Results in Retinitis Pigmentosa**

In this section, the classical RP symptoms and how they manifest when imaged by medical imaging techniques for diagnosis will be discussed. Patients with RP generally first notice macroscopic deteriorations in their visual acuity, experiencing primarily nyctalopia or night blindness, and secondarily peripheral vision loss (14). Both of these vision deteriorations are a consequence of photoreceptor rod cell death: rod cells are known to function in scotopic vision of low light levels, thus their damage and death would be known to cause nyctalopia; rod cells also heavily populate the peripheral retina, with sparse concentrations in the central retina, meaning their loss would most heavily affect peripheral vision (38). Patients in the later disease stages will start to experience a further narrowing of the field of vision in concentric rings, as rod cell death triggers retinal layer secondary changes, which catalyze cone cell damage and death (14). As it's known that cones are most tightly packed within the central retina and are scarce within the peripheral retina and rod cell population is the inverse of this situation (38), the sequential rod cell death death circularly shrinks the peripheral field of vision, often causing mid-peripheral blind spots or scotomas (15), to central vision; then, the cone death triggered by the rod dysfunction further narrows the central field of vision (14). In the pathology's end-stage, the central field of vision is completely lost (14). The exact mechanism in which secondary cone death is triggered by rod cell death is still poorly understood, however postulations involving rod death endotoxin release triggering cone death, superfluous oxidative cone stress due to the loss of rod metabolism, and the absence of secreted rod survival factor preventing cone survival have been presented (69). Additional ophthalmic derelictions may also occur: approximately half of RP patients are afflicted with posterior subcapsular cataracts (PSC), opaque clumps of lens fibres on the lens' posterior subcapsular area, or lens back/posterior which occlude vision; many also experience bright flashes of flickering light or shapes in the absence of external light stimulus, otherwise referred to as photopsia (14). A small subset of patients will also present with cystoid macular edema (8), (14) (swelling of the macula or central retina (38) due to fluid retention in quasi-cystic organizations), macular holes, the formation of epiretinal membranes (ERMs), thin scar tissue layered across the macula otherwise known as macular puckers (14), bone spicule pigmentation (RPE cell disintegration causing pigment to migrate into the retina's interstitial cavities) (8), and fundus hyperfluorescent rings (16); these abnormal physiologies within the macula further decrease the patient's ability to visually construct detailed images (14). Certain cases of RP are also known to be syndromic, meaning the pathology affects tissues and organs other than the retina; in contrast, the non-syndromic RP cases are disease manifestations in which the pathology affects only the retina (14). Syndromic cases are most commonly caused by autosomal recessive gene mutations, such as those of autosomal recessive Usher syndrome, and Bardet Biedl syndrome (14). Some of the more severe symptoms which syndromic RP patients may experience include learning disabilities, abdominal obesity, hearing loss, renal defects, diabetes and other endocrine disorders, and ataxia, or difficulty with coordination, balance, and speech (14). In extreme cases, patients develop simultaneous exudative vasculopathy, a build-up of plaque in the foveal region due to a breakdown of the blood-brain barrier, which can even manifest as retinal vascuproliferative tumours (30).

These macroscopic symptoms of worsened visual acuity and decreased visual field patients may experience are generally easily detected through physical evaluation. As well, these deteriorations in visual function such as degradations in visual acuity (VA) and contrast sensitivity often occur in later disease stages, thus they do not serve as informative parameters for visual function, especially in early disease stages (37). For true diagnosis, medical imaging techniques, sometimes in combination with genetic testing, must be implemented, as the former would confirm the microscopic symptoms which accompany the macroscopic physical onset, and the latter would confirm the patient's specific genetic RP-causing mutation (11), (14). Those subtler symptoms include bone spicule pigmentation (7), waxy optic disc pallor, decreased retinal electrical responses as quantified by lower electroretinogram (ERG) amplitudes, and a diverse array of morphological changes that occur in the retina pigment epithelium (RPE) (13). Interestingly, each imaging technique can reveal characteristic symptoms of RP undetectable or otherwise difficult to ascertain from other techniques. The diagnostic imaging and results of each medical imaging technique to be discussed will be traced in the sequential order of the techniques elucidated in the Background section, though it should be noted a significant amount of research teams used a multimodal approach to imaging, thus there will be some overlap between the various imaging modalities. Starting with adaptive optics AO, the RP imaging results will be discussed in two portions, those attained with confocal scanning systems, and those attained with flood-illuminated systems. As well, since certain AO results were compared and correlated with OCT findings, these findings will also be presented in the context of how they relate to the AO findings. As mentioned above, the most common implementation of scanning confocal AO systems is scanning laser ophthalmoscopy (SLO), otherwise known as AOSLO (52). Generally with AOSLO, the target metrics measured are cone density, cone spacing, Voronoi analysis for mosaic geometry assessment in which adjacent cells are counted relative to their separation distances, reflectivity, and cone directional dependence (or anisotropy) for some given retinal eccentricity (52). It's well known that cone density decreases (meaning inter-cone spacing increases) with RP, as was demonstrated by the findings of Song et.al; Song's team used split detection AOSLO (SD-AOLSO), in which successive imaging frames are subtracted to partition the detection of the leftward and rightward light relative to the aperture, and confocal AOSLO (cAOSLO) to investigate the foveal cone density in two autosomal recessive RP patients, one with non-syndromic RP, the other with syndromic RP, more specifically Usher syndrome (52). It was found that there existed a high degree of redundancy in the foveal cone population, as as much as 38% of cone density was lost before an appreciable decrease in visual acuity was detected, morphological changes which remained undetectable through OCT (52). Sun et.al further investigated Usher syndrome with SD-AOSLO and cAOSLO to reveal that syndromic cases of RP demonstrate greater foveal and parafoveal cone density loss relative to the losses seen in non-syndromic RP; moreover once again, these density changes weren't evident in OCT images, as the retinal interdigitation zone (IZ), otherwise known as the cone outer segment tips (COST) and rod outer segment tips (ROST), as well as the ellipsoid zone (EZ), or the inner segment-outer segment (IS/OS) zone, revealed no morphological abnormalities (52), as supported by the work of Mitamura et.al (22). Sun's group posited this lack of visible changes in the OCT scans was due to the fact that different RP cases have different localized effects within the retina, with these Usher syndrome RP patients possessing marked decreases in the amount of normal waveguiding cones (52). As already established, the redundancy in the cone population gives way to the possibility of dramatic losses in cone density before any noticeable deteriorations in visual acuity manifest; thus, early detection of photoreceptor cell loss is essential for the early diagnosis of RP (53). Nakatake et. al's group endeavoured to do just that, using AOSLO to reveal the parafoveal cone density changes in early stage RP while comparing them to OCT findings; they found that RP patient's mean cone densities were lower at foveal eccentricities of 1.0mm relative to the control patients, even in patients with normal interdigitation (IZ) zones as seen in OCT findings and normal visual sensitivities as seen in static perimetry (microperimetry) tests (53).

Moving onto the RP patients imaged with flood-illuminated adaptive optics, Gale et.al's research team made notable contributions in two instances, one in 2015, and one in 2019; the earlier rendition of Gale's team strove to identify correlations between visual features in RP patients imaged with a variety of methods: flood illuminated AO, FAF (fundus autofluorescence), and SD-OCT (spectral domain optical coherence tomography) (54). They found there existed high correlation between AO cone reflectivity profile loss, FAF hyper-autofluorescence changes, and OCT IZ zone abnormalities (54); note in this case, IZ morphological changes were detected in contrast to the results obtained above with AOSLO in which no noticeable morphological abnormalities were observed in the OCT images (52), (53), illustrating once again the diversity in which RP can manifest in patients. Not only did there exist correlations between the three parameters, but also the exact phase of cone photoreceptor degeneration could be surmised with the correlations found with FAF and SD-OCT, as the type of cone reflectivity AO profile—healthy cone mosaics, hyper-reflective blurred quasi-cones, disorganized higher frequency hyper reflective patches, and hypo-reflective spots of lower frequency—correlated to the degree of degeneration (54). Gale's team interestingly, through their observation and classification of the various AO cone reflectivity profiles in the degenerative foveal areas, observed ambiguity in how a cone can be structurally defined; in other words, what exactly comprised a cone remained slightly nebulous (54). After the first submission of flood-illuminated AO imaged RP, Gale's team set out on a second endeavour in 2019, using flood-illuminated AO once again for RP cone parametric measurement, but this time with the aim to classify the repeatability of measurements (55). Gale et. al's second work is significant, as at the point of publication, few repeatability studies of cone density in RP patients with flood-illuminated AO had been conducted (55). Gale's team compared imaging results from 10 different RP patients with various pathological genetic mutations and phenotypes—poor visual acuity, decreased visual field, history of cataracts, OCT-detected cystoid macular edema—with images obtained from 11 healthy volunteers, and with the control group (55). 25 overlapping $4°$ x $4°$ flood-illuminated images were captured, then partitioned into equally spaced regions of interest (ROIs) for more facile analysis with a MATLAB algorithm for rod and cone identification which the authors customized for this study (55). A sample of the image's ROIs at various eccentricities, in which the image quality was deemed either "good" or "poor" unanimously by three subjective graders, was analyzed for cone density, cone spacing, coefficient of variation (CoV), coefficient of repeatability, repeatability, and a novel parameter introduced by the authors, cone location similarly (CLS) between repeated imaging sessions (55). CLS quantifies the common or similar cones between two images of the same location; the results were consistent with previous results, as well as those within the field: average cone density was found to be significantly reduced in RP patients compared to control patients (55). The authors also found the standard deviation (SD) and coefficient of variation (CoV) of cone spacing to be lower in high-quality images compared to those of low-quality images due to the fact that better image quality reveals clearer visualization of the mosaic; it should be noted in healthy cone mosaics, both the SD and CoV of cone spacing should approach zero as healthy cone mosaics are highly ordered within a neatly packed, hexagonal matrix for maximal packing efficiency, thus their cone spacing should be uniform across the entire mosaic (55). The CLS was found to be higher in good quality images compared to poor quality images, however this observation can't be fully attributed to the fact that RP patients have lower CLS from poor cone organization, cone mosaics, and wave-guiding abilities alone; confounding factors that affect cone reflectivity and CLS across subsequent images of the same location include the Stiles-Crawford effect, an anisotropy wherein higher photoreceptor responses arise from light originating from the central pupil compared to the responses due to light from the pupil edge as a result of cone's wave-guiding properties, cone outer segment renewal, diurnal variations, and cone alignment (55). Regardless, though retinal location and spatial orientation remain unchanged across subsequent imaging sessions, RP patients demonstrated higher reflectivity variability and reflectivity artifacts than those of the control group, a quality which is characteristic of retinal disease and ultimately retinal death (55). Thus, Gale et.al found across two experimental endeavours that though the cone density repeatability metrics showed no significant difference in RP patients relative to the control patients, RP patients did demonstrate decreased cone density and higher variance in cone reflectivity profiles which could be correlated with FAF and SD-OCT findings, both consistent with the expected finding for RP (54), (55).

Almost all of the above AO imaging results discussed involved some comparison to other images obtained with different modalities, most commonly to those obtained with OCT, either for metric correlation or subjective comparison. Consequently, it logically follows to speak on OCT RP diagnostic and imaging results, as OCT has already revealed that the IZ morphological abnormalities expected in RP patients is sometimes but not always observed (52), (54), (55). Because OCT is such a universal technique in retinal imaging, results in which authors didn't exclusively use OCT, but rather included other imaging modalities for comparison, are numerous; this subsection will instead focus on those results in which OCT was the primarily used imaging technique. These OCT results will include those obtained with conventional TD-OCT, SD-OCT, and OCTA. In terms of retinal imaging, OCT allows for high resolution imaging of retinal morphology, wherein each highly reflective layer in the retina is clearly distinct: the choroidal vessels, Bruch's membrane, the RPE, IZ, ellipsoid zone (EZ), external limiting membrane (ELM), otherwise known as the OLM as discussed above, outer plexiform layer (OPL), inner plexiform layer (IPL), and retinal nerve fibre layer (RNFL); note these reflective layers partition the retina in a similar but not identical manner to which the layers are defined in the Background section of normal retinal physiology (7). Generally in OCT retinal imaging, three distinct, highly reflective lines can be visualized for photoreceptor evaluation: the ELM, EZ, and IZ (7). Figure 14 demonstrates the normal retinal physiology as imaged by SD-OCT, with the ELM (number 9 on Figure 14) visualized as the topmost, faintest reflective line, the EZ (number 11 on Figure 14) visualized as the central, intermediately reflective line, and the IZ (number 13 on Figure 14) visualized as the bottommost, intensely reflective line.

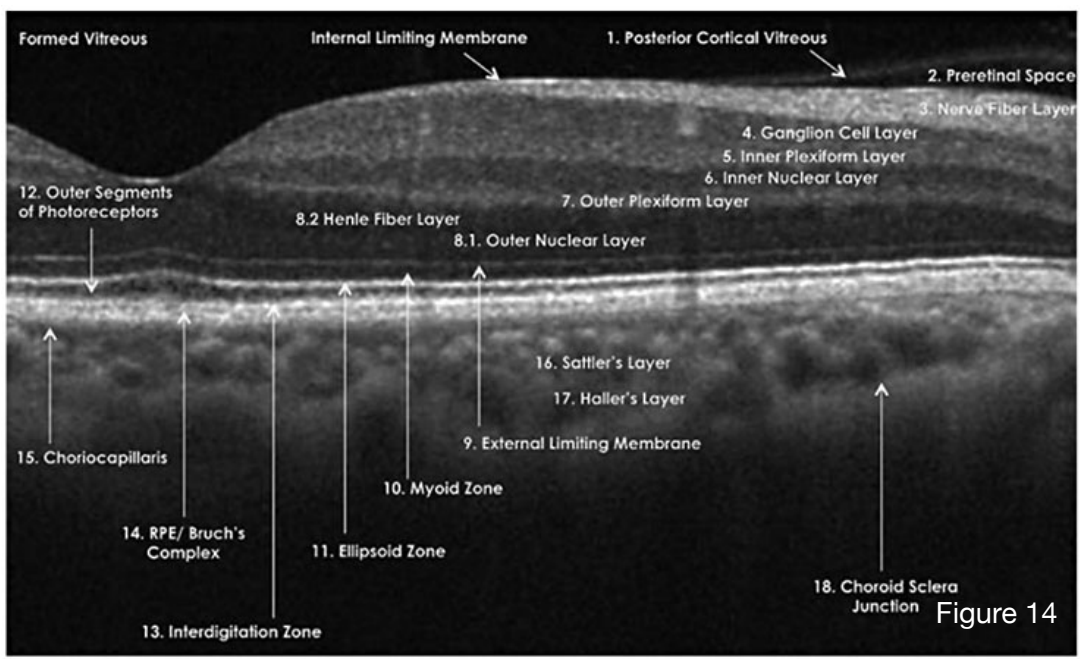

Figure 14

RP's earliest detectable morphological abnormalities occur in the photoreceptors, specifically in the cell's outer segments presenting as disorganization within the OLM, EZ, and IZ bands (7). Tamaki et.al measured each of the three bands in RP patients using SD-OCT and found that the length of the bands had morphed to a disorganized, pathological state in which all bands shortened relative to the lengths of healthy patient bands: the OLM band was found to be the longest, the IZ band found to be the shortest, and the EZ band's length existing as the intermediate (7). These length alterations imply RP photoreceptor disorder and degradation occurs first as a shortening of the outer segments, as its corresponding IZ band was the most shortened; the degradation then continues along the length of the photoreceptor, reaching the EZ, then the OLM (7); this precise sequential photoreceptor deterioration is further supported by the results of Mitamura et.al (22). The fact that the shortened IZ band is the first detectable morphological abnormality in RP patients using OCT explains why IZ band abnormalities were only detected in some of the above AO experiments, as other morphological changes, for example cone density deteriorations, can occur even earlier than photoreceptor shortening in the disease progression and remain undetectable by OCT. In patients in more advanced stages of RP, ONL thinning, or the thinning of photoreceptor nuclei, was also detected via OCT; it's been found by a vast collection of researchers that specific OCT-facilitated measurements of retinal morphology can provide quantitative markers for determining disease progression (7). For example, Rangaswamy et.al found that photoreceptor layer thickness (in both the IZ and ONL) is directly proportional to the visual field, with the EZ line length specifically serving as a landmark for the visual field boarder; thus, the detection of an in tact EZ line corresponded to relatively conserved photoreceptor function (7). Further correlations between retinal structures and visual function were investigated by Yoon et. al, who determined that the preserved photoreceptor area, calculated either from the en face images themselves, or from the multiplication of the horizontal and vertical EZ lengths, was shown to correlate with the visual field area, although with a weaker correlation than that of the EZ line and visual field area (7). Sugita et.al strove to additionally investigate EZ length correlation with mfERG amplitude to find the relation between EZ line length and macular function; though longer EZ lengths were found to be correlated with higher mfERG amplitudes, the correlation was weak at best due to pathological alterations affecting visual function that remain undetectable by OCT images, as previously mentioned, and the IZ line itself affecting the mfERG amplitude, not the EZ line (7).

Moving to the RP morphological changes in the inner retina as detectable by OCT, there exists a notable asymmetry in the thinning of the ONL, inner nuclear layer or INL (containing the bipolar, horizontal, and amacrine cells), and the ganglion cell layer (GCL), as illustrated by the work of Aleman et. al and Hood et. al, both of whom found that the width of the INL and GCL was unaffected despite continual thinning of the ONL (7). Aleman's team extended the retinal thickness disparity to find that significant correlation existed between outer retinal thinning and preserved inner retinal layers in RP, wherein thinning of the ONL often doesn't correspond to thinning of the overall retina due to the simultaneous, compensatory thickening of the inner retina (7). The thickening of the inner retina was postulated by Aleman et. al to be due to the neural-glial restructure of bipolar dendrites pulling back and the amacrine and Müller cells becoming more reactive, biochemical and physiological changes which occurred in response to outer retinal thickening (7). Curcio et.al and Beltran et.al proposed further mechanisms for ONL thickening and INL thickening, being that the Henle fibres, the Müller cells and photoreceptor axons comprising the INL, were resistant to RP-induced morphologies, or the retinal nerve fibre layer (RNFL) swelled in response to RNFL loss respectively (7). Though there exists no universally agreed upon consensus of RP GCL changes, correlation between the two have been found, indicating the GCL informs on the additional RP-caused inner retinal morphological abnormalities; Vámos et.al demonstrated that RP patients with thicker, more in tact GCL and IPL layers possessed detectable mfERG signals, while the mfERG signal of patients with thinner GCL and IPL layers remained undetectable (7).

Additional OCT-detected RP-induced morphological changes occur in the nuclear layers, RNFL, macula, and choroidal vessel layer. First, hyperreflective foci (HF)—migrating RPE cells, macrophages, and vessel-excreted lipoproteins which appear as dots with equal if not greater reflectivity than that of the RPE in SD-OCT images—have been detected in RP patients in both the inner and outer nuclear layers of the photoreceptors, as well as the subretinal space by Kuroda et.al (7). Kuroda's team used SD-OCT to observe the difference in HF presence across RP patients of various disease advancement stages, finding that ONL HFs due to photoreceptor and RPE cell death were accompanied by discontinuous, pathologically malformed RPE and EZ lines and decreased best corrected visual acuity (BCVA) when compared to RP patients without ONL HF (7). Gupta et.al, Marc et.al, and Li et.al additionally found that HF aren't sequestered to the ONL of the nuclear layer; the three teams all found HF within the INL in RP patients, localized to the pathologically deteriorated EZ-RPE boarder (7). In contrast to the HF within the ONL indicating photoreceptor death, all three teams concluded INL HFs were simply artefacts of normal retinal function (7). Second, looking now at the RP-induced changes in the RNFL, it was found that though ganglion cell death continued to progress linearly with disease progression, the RNFL doesn't thin uniformly in an atrophic pattern but rather both thins and thickens (7). Fibrous astrocytes were hypothesized by Szamier et.al to cause RNFL thickening; as these astrocytes usually occur on the optic nerve head surface, the optic disc's most proximal RNFL, the thickest of the nerve fibre layers, causing the characteristic RP waxy optic disc pallor (7), (14). Though the exact locations and extent of RP RNFL thickening and thinning varies across findings within the field, Oishi et.al demonstrated that late stage RP patients of older ages characteristically showed thinned RNFL when OCT-imaged (7). As well, the radial peripapillary capillary network was found to decrease in density in RP patients, the density of which correlates with RNFL thickness, supporting why both metrics reduce across more advanced stages of RP (9). Regardless, given the inconsistency of RNFL measurements, GCL thickness is a better diagnostic and disease monitoring metric for RNFL than RNFL thickness itself. Third, macular volume and thickness, metrics of determining macular function due to blurring of retinal layers in late stage RP, tend to decrease in RP patients, specifically in the integrity of the photoreceptor line (8), and is directly proportional to mfERG amplitude, as illustrated by Tamaki et.al and Vámos et.al respectively (7). Tamaki's team also found a positive correlation between macular thickness and BCVA (7). The number of macular abnormalities, which were if not as effectively, but rather more effectively imaged with OCT than other fluorescent imaging techniques such as fluorescence angiography, were higher in RP patients relative to those of unaffected subjects, with cystoid spaces, ERM, macular edema, vitreomacular traction (VMT) syndrome wherein the vitreous incompletely detaches from the macula, macular holes (7), and cystoid macular edema (8) as was mentioned at the start of this section. Finally, RP-associated changes in the choroid vessel layer were detected via OCT (7). The migration of blood vessel-proximal RPE cells induces extracellular matrix (ECM) deposition within RP patients; as the matrix progressively thickens and encroaches on the vascular lumen, blood flow is continually obscured (9). Thus, the choriocapillaris (the innermost choroid responsible for RPE and photoreceptor nourishment) atrophies due to RPE and photoreceptor cell death, abnormal choroidal structures such as scalloped choroid and sclera borders form, and the large choroidal vessel layer thins (8), all three of which deteriorate BCVA. Adhi et.al additionally reported a direct proportionality between sub-foveal choroidal thickness and central retinal thickness, though generally, choroidal thickness doesn't serve as a definitive diagnostic RP metric given the inhomogeneity in sub-foveal choroidal thickness correlation results with RP severity or duration (7).

Moving now to the RP retinal symptoms as imaged by OCTA, researchers most commonly investigate perfusion density (total perfused vasculature area per measured unit area) changes in foveal structures, EZ line width, and choriocapillaris blood flow (CBF), to be expected as those parameters are the ones OCT-A is most suited to quantify given its ease in visualizing retinal and choroidal vasculature as shown by the work of Ruben et.al (9), and Takagi et.al (10). Ruben et.al imaged 28 RP patients over a period of 1-2 years to quantify RP-caused retinal vasculature changes, and correlate them with EZ line width, a parameter already demonstrated above to be directly proportional to visual function and acuity, specifically BCVA (9). Moreover, Ruben's team wanted to demonstrate the usefulness of perfusion density and the area of the foveal avascular zone (FAZ) as metrics for RP diagnosis and monitoring (9). The team quantitatively measured perfusion density in both segments of the capillary plexus, then used these perfusion density rates to extrapolate RP progression rate; they found that in RP patients, both perfusion density and FAZ area changed annually in the SCP and DCP: perfusion density decreased while FAZ area increased (9). Ruben's team discovered that despite the decreased perfusion and and increased avascularity within the fovea, choriocapillaris blood flow didn't change significantly over the two year data collection period (9). Contrast the consistency in choroidal blood flow with the known decreased retinal blood flow velocities in RP patients as discovered by OCTA. Though choroidal blood flow showed no observed change, significant differences between the quantified blood flow of the SCP versus the DCP were found; the SCP demonstrated healthier blood flow low relative to the DCP, with both a higher perfusion density and lower avascularity with a smaller FAZ (9). Regardless of the size comparison between the FAZs of the SCP and DCP, decreased flow area within the capillary plexus in RP patients was a result agreed upon by Tagaki et.al (10). Correlations between metrics of visual function and SCP blood flow parameters were also found: perfusion density was strongly associated with EZ line width, and FAZ area was strongly associated with BCVA; again, choriocapillaris blood flow showed no correlation with any visual function parameters (9), a consistent results across both Ruben's and Tagaki's work. These findings seem to suggest in the capillary plexus, the SCP's structure and function is preserved despite RP disease progression as its preservation ensures visual function is maintained. Tagaki's team found further flow area changes, primarily imaging and measuring them using OCTA, but with the additional incorporation of SD-OCT, wide-field FAF, and microperimetry via the Goldmann perimeter as supplemental imaging techniques (10). SD-OCT was used to image and measure the inner segment ellipsoid (ISe) and OLM, whereas wide-field FAF was used to identify areas without abnormal fluorescence (the normal FAF area ratio), and the Goldman perimeter was used to evaluate the visual field (10). Tagaki's team imaged RP patients with good VA to show that once again, despite healthy visual acuity and sensitivity, subtle deteriorations can occur before any visual function or morphological changes can be OCT detected, in this case; those deteriorations were of overall vascularity (10). A unique finding was that flow area was strongly associated with ISe and OLM lengths only in the SCP, not in the DCP; and, no correlation between flow area and the FAF area ratio was found (10). Jauregui et.al also used the same three imaging methods, SD-OCT, OCTA, and FAF, to monitor disease progression of a case of autosomal dominant RP, specifically caused by a mutation in the RPE65 gene; the pathological mutation was confirmed via peripheral blood cell sequencing (12).

DNA

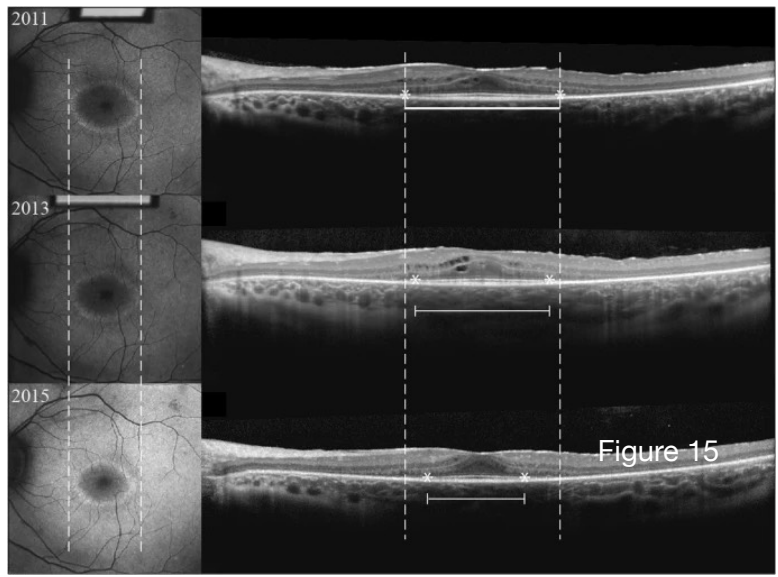

Figure 15

As demonstrated by the work of Tagaki et.al, and Jarequi et.al, fundus autofluorescence is often paired with OCT for more informed imaging and disease diagnosis; the common implementation of the OCT-FAF duo for RP imaging is exemplified by the work of Cabral et.al, Ueno et.al, Andersen et.al, and Dysli et.al (15), (16), (18), (19). The diameter of the fundus hyperfluorescent ring and fluorescent lifetimes are two commonly measured metrics between the four research teams (15), (16), (18), (19), with all groups but Dysli's SD-OCT-quantifying the EZ either as the EZ line width or EZ band diameter, parameters highly associated with BCVA (15), (18), (19). Cabra's team imaged 81 RP patients using SD-OCT and SW-AF with an average follow up time of 3.1 years (15), with patients representing all three Mendelian inheritance patterns, with some having syndromic Usher syndrome, and others cystoid macular edema (CME), which as an aside, distorts the appearance of the EZ line in SD-OCT (15). The researchers found results consistent with those obtained above (22): the average EZ line width, as seen in the SD-OCT images on the right side of Figure 15, shortened considerably, corresponding to a loss of visual field of about $0.5°$ per annum, and the average hyperfluorescent ring diameter, as seen in the SW-AF images on the left side of Figure 15, constricted considerably in both the vertical and horizontal direction over the data collection figure period; all three values shortened at a rate of approximately 140 µm per year (15). The disease progression rate as quantified by the rate of EZ line width shortening and ring diameter constricting was found to be proportional to the starting EZ line width and ring diameter, implying the more advanced the stage of disease, the smaller the starting EZ lines and rings, the slower the progression (15). This slowing of disease progression as the disease stage becomes more advanced is characteristic of RP, and mathematically described as degenerative exponential decay (15). Cabra's team further found the disease progression to be asymmetric in about 20% of the patients between the right and left eye (15); thus, they demonstrated not just the exponential decay of RP degeneration, but also the fact that EZ line width and hyperfluorescent ring diameter can serve not just as diagnostic metrics, but also metrics for estimating the current stage of disease. The second team Ueno et.al used FAF, OCT, and AO fundus imaging to examine nine RP patients, finding that macular degeneration occurs early in the diseases stage; the exact area within the central macula wherein photoreceptors were preserved was decided from the OCT and FAF images, as hyper-fluorescence in the former, and normal outer retinal structure in the latter (16). AO fundus imaging also identified the preserved macular regions, correlating with the surviving photoreceptors which were demarcated in FAF images by hypo-reflective clumps of melanin (16) and shrunk over the collection period of four years; these findings seem to that the degenerative area is flanked by the inner borders of hyper-fluorescence.

Figure 16 data imply FAF lifetimes

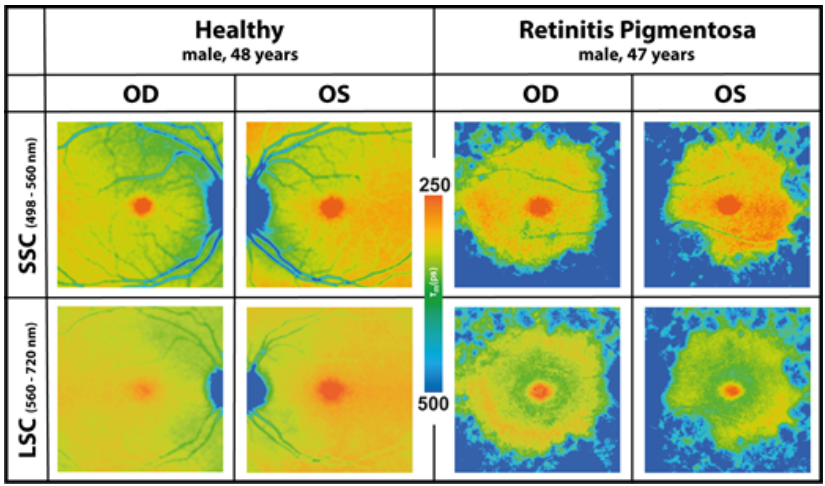

The third team, Andersen et.al, also found results consistent with shrinking fluorescent rings and the rings bordering the degenerative zone (18). The researchers used FAF and SD-OCT for RP imaging of 33 patients with all three genetic RP types, but they focused on the quantification of fluorescence lifetimes in addition to the measurement of ring diameter in different areas of the macula; Andersen's team implemented fluorescence lifetime imaging ophthalmoscopy (FLIO) in both short spectral channels which detected shorter wavelengths (SSC) and long spectral channels (LSC) which detected longer wavelengths (18). As mentioned in the background section, FLIO allows for the collection of quantifiable information on the current disease stage of RP in a patient, even in early stage RP cases, as metabolic changes often occur before the visible physical deteriorations occur in later stages (18). FAF lifetimes were founds to be significantly higher in RP patients versus those of the control group, especially in the outer macula, consistent with the atrophic regions as seen in Figure 16 (18). Andersen's group also noticed ring-like patterns, most noticeably in the peripheral macula, with FAF lifetimes less than those in the atrophic outer macula, but greater than those in the general macular area (18). Prolonged FAF lifetimes in the periphery, specifically the outer macula, which ring the fovea were found, characteristic of RP patients and correlated with the atrophic retinal regions as SD-OCT imaged due to photoreceptor degradation and phagocytosis; fluorescent ring size is also known to be directly proportional to BCVA and retinal sensitivity as demonstrated by EZ line SD-OCT measurement (18). Though the ring pattern is known to border the degenerative area across all RP patients (22), the specific FAF patterns varied across the patients, with autosomal dominant and Usher syndrome patients demonstrating the most pronounced rings, and autosomal recessive patients demonstrating less pronounced rings and the lowest FAF lifetimes (18). The approach and results of Andersen's team were both similar to those used by Dysli et.al, wherein FLIO FAF and SD-OCT were employed for fluorescent lifetime analysis in 43 RP patients (19). Once again, two spectral channels, one for short wavelengths and one for long wavelengths, were used in FLIO, and SD-OCT and BCVA data were collected for correlation with the FAF data (19). FAF lifetimes were found to be positively correlated with photoreceptor atrophy and manifest as hyper-fluorescent perifoveal rings wider than those within the fundus, consistent with the above; the longest lifetimes correlated to total RPE and photoreceptor atrophy (19). OCT findings of preserved retinal structure, EZ band diameter, and BCVA were inversely proportional to the FLIO lifetimes, as to be expected as prolonged FAF lifetimes in RP patients are known to correlate with increased photoreceptor degradation and visual function loss (19). Thus, the above four teams exemplified that FAF and FLIO imaging provide consistent and informative methods of diagnosing and monitoring RP, especially when paired with other imaging techniques.

Shifting now to the diagnostic monitoring via colour Doppler flow imaging (CDFI), this technique is less widely used for imaging RP relative to other vasculature imaging techniques such as OCTA, as OCTA allows for microvascular and macular thickness changes to be imaged and detected prior to the later disease stage large-scale vascular changes which occur and can be detected by CDFI (20). Regardless, notable contributions to CDFI RP imaging from Wang et.al (20) and Cellini et.al (59) are worth mentioning. With Wang's work, it's no surprise that OCTA was used to supplement CDFI, given OCT's prevalence in ocular imaging and the ease with which vasculature is visualized via OCTA (20). Contrary to the above SD-OCT results which demonstrated no statistically significant change in ophthalmic blood flow, Wang et.al operated off the phenomena that both retinal and choroidal flow are impaired in RP due to elevated endothelin-1 (ET-1) blood concentration which is known to increase vasoconstriction (20). The researcher's main objective was to determine if retinal large vessel changes and terminal vessel changes were related, and to compare CDFI and OCTA in terms of which method more effectively imaged RP vascular changes, both within the superficial vascular layer (20). The RP patients were diagnosed on ophthalmological exam, with patients presenting with night blindness, abnormal fundus visualization, visual field loss, and strange ERG scans, and grouped by visual acuity into either the high- vision group or the low-vision group; additional ocular examination included slit-lamp biomicroscopy and the measurement of both intraocular pressure (IOP) and visual acuity (VA) (20). The parameters of interest included vascular area density (VAD) between the GCL and IPL, with VAD defined as the total vasculature area divided by total image area, fovea avascular zone (FAZ) (measured once again as was seen in the OCT-FAF results), and retinal layer width from the ILM to the RPE (20). Coefficients of variation (CV) in both the CDFI and OCTA results were also calculated for statistical repeatability and reproducibility (20). All measured CDFI and OCTA parameters decreased significantly in RP patients, with the low-vision group demonstrating more severe decreases in the FAZ area,

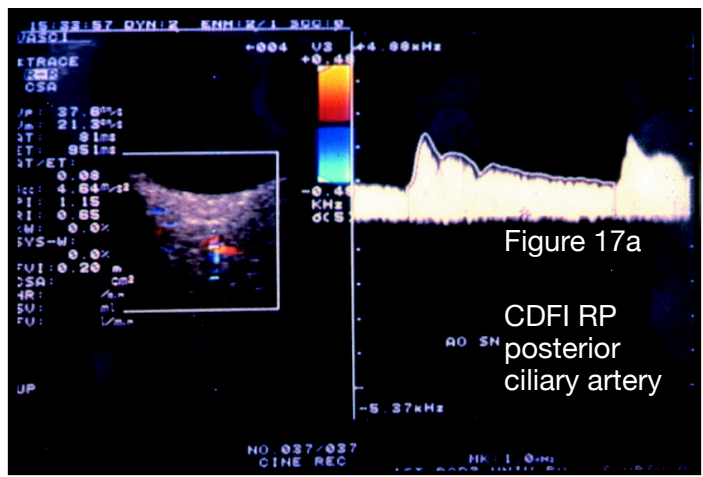

Figure 17a

CDFI RP posterior ciliary artery

foveal VAD, parafoveal nasal VAD, and overall retinal thickness relative to the low-vision group (20). Though no statistically significant difference between the CDFI parameters measured in the low-vision and high-vision group was observed, CDFI parameters of peak systolic velocity (PSV) and time-averaged maximum velocity (TAMX) were found to correlate greatly with vision (20). As expected, the OCTA derived FAZ size, VAD in both the foveal and nasal side, and retinal thickness were also strongly correlated with vision, and the central retinal artery PSV and TAMX correlated intensely with OCTA superficial VAD (20). Cellini et.al further explored the correlation between ET-1 plasma levels and blood flow in RP patients as mentioned above, measuring parameters of both using CDFI and laser Doppler flowmetry (59). 20 patients were RP diagnosed due to their decreased visual fields, abnormal ERG scans, retinal bone spicule pigmentation, attenuated blood vessels, and waxy-coloured optic nerve heads which presented upon ophthalmic examination; patients with posterior subcapsular cataracts (PSCs) and macular cystoid edema (MCE), as well as those presenting with other general or systematic disorders were rejected form the study (59). Not surprisingly, Cellini's team found that blood flow was significantly reduced in RP patients relative to that of healthy controls, with the CDFI-obtained peak stroke volume (PSV) of the RP patients decreasing by 15% and 39.3% in the ophthalmic and posterior ciliary arteries (Figure 17b) respectively compared to that of healthy controls (Figure 17a); this two-fold PSV decrease was found to be correlated with the elevated ET-1 plasma levels discovered in RP patients, wherein levels were 38.2% higher in pathologically afflicted subjects (59). Laser Doppler flowmetry results further exemplified this blood flow decrease, with RP patients presenting a relative decrease of 57.3% and 9.8% in baseline peak flow (PF) and capillary blood flow respectively; the cold pressor tests for cutaneous capillary perfusion as measured by laser Doppler flowmetry resulted in increased maximal cold-induction flow reduction times and recovery times (59).

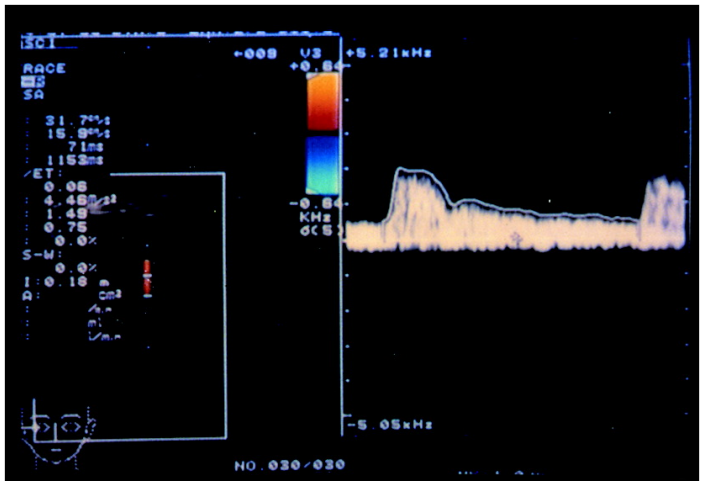

The second to last diagnostic imaging technique results to be discussed will be those derived from microperimetry (MP). Though MP is more often than not performed in tandem with other imaging techniques such as SD-OCT and FAF (21), significant results derived from experimental setups involving MP as the only imaging technique will also be presented (36), (37). Given the understanding that retinal sensitivity, the target metric of MP, informs greatly on overall retinal function, researchers often employ MP as a means of observing pathological progression (37). The first research team, Iftikhar et.al, imaged 106 RP patients during a 5 year data collection period using MP for paracentral, mean, and central sensitivity measurement, paired with the classical FAF, for hyper-fluorescent ring measurement, and SD-OCT, for EZ line width measurement, to monitor disease progression (21). Iftikhar's team found that MP-derived mean and central sensitivity were directly proportional to visual acuity; in fact, statistically significant correlations between all measured parameters were found excluding paracentral sensitivity (21). Consistent with other results, the results of Iftikhar et.al also demonstrated that visual acuity, ring area and diameter, and all three retinal sensitivities decreased significantly per annum (21). These results were exactly mirrored by those obtained by Buckley et.al, who used MP to image and monitor RP cases derived from RPGR (retinitis pigmentosa GTPase regulator gene) mutations (36). Buckley's team demonstrated the usefulness of MP as an imaging system to monitor the outcome of potential treatments, which will be discussed in the subsequent section (36). As RPGR is such a common case of RP, especially in X-linked cases, it's no surprise the final team, Anikina et.al, also investigated this specific gene variant in 76 subjects through MP imaging over the course of two years (37). Anikina's team strove to not just investigate the efficacy of MP as a means of RP imaging, but also to statistically test for the repeatability and intraclass correlation of the results, specifically of the mean sensitivity (MS); analysis of volume and topography was also performed to obtain BCVA, contrast sensitivity (CS), MS, total volume (V_tot), and central 3-degree field volume (V3) (37). MS, V_tot, and V3 were all found to have both high ICC values (implying a high degree of retest-ability) and intensely positive correlations between the three parameters (37). This high correlation implies there exists a high degree of interocular correlation between the target parameters; though it should be mentioned, MS and V_tot interocular progression rates appeared to be largely symmetric in both eyes, whereas the V3 progression rate demonstrated a noticeable ocular asymmetry (37). Anikina's team also discovered retinal sensitivity to abruptly and severely degrade starting in the patient's 20s, and progressing into their 30s of age, with the majority of patients demonstrating dramatic decreases in MS, V3, and V_tot (37). The Spearman correlation coefficient was also used to calculate a strong direct proportionality between BCVA and the four metrics of CS, MS, V_tot, and V3; direct proportionality between CS and the three metrics of MS, V_tot, and V3 were also found (37). In fact, BCVA and CS were found to have the highest correlation, followed by that between BCVA and V3, and CS and V3; given that statistically insignificant correlation between the progression rates of BCVA and CS and MS and V_tot were found, the team concluded BCVA, CS, and V3 serve as more informative metrics for identifying disease stage and monitoring disease progression, especially in light of central macular function only starting to worsen in RP's end stages (37).

With the succinct summary of MP-derived RP imaging results concluded above, the final RP imaging technique results will be covered: those obtained through MRI. Though MRI is a slightly unconventional ocular imaging technique and is usually applied to larger scale organs and tissues, it is exactly this unconventionality that lends to the unique results and fascinating insights into RP diagnosis and disease monitoring obtained through MRI imaging. With any novel imaging application, MRI was often first implemented in rat and mouse models before being used for human imaging for example by teams such as Zhang et.al (35), Li et.al (57), and Muir et.al (58). Interestingly, all three teams investigated retinal and choroidal blood flow in RP subjects and operated off the fact that retinal blood flow was greatly affected by RP, most generally being reduced or restricted as the disease progressed, implying MRI is more suited to image vasculature in the context of RP (35), (57), (58). This implication is further supported by the fact that MRI, though possessing a poorer spatial resolution than other optical imaging techniques, does in fact allow for tissue perfusion and blood quantification measurement without individual vessel visualization and a high depth resolution and wide field of view (35). As well, though OCT allows for clear visualization of retinal architecture, it is a method from which physiological parameters are difficult to deduce (35). In spite of the advantages of MRI over OCT for RP imaging, there exists a trade off between the high depth resolution of MRI, and its long data collection period which exposes its susceptibility to motion artifacts (57). Keeping these consideration in mind, it follows why each team chose to MRI image blood flow. The first group led by Li used the Royal College of Surgeons (RCS) rat, an animal model with MERTK mutation derived RP; the MERTK mutation specifically impedes photoreceptor segment phagocytosis, resulting in shedding and accumulation of the photoreceptor outer segments and spontaneous photoreceptors deterioration (57). Li's team used MRI to image and quantify the RCS rat's choroidal and retinal blood flow, as well as resolve and measure retinal layer thickness at various RP disease stages, allowing for precisely depth resolved imaging which conventional optical imaging techniques, excluding OCT, can't achieve (57). The researchers employed continuous arterial spin labeling (cASL), wherein carotid blood flow is magnetically labelled in one image, then unlabelled in the next, to quantify blood tissue perfusion from the differences between subsequent images; static signals were also suppressed with the implementation of an inversion recovery pulse, improving sensitivity (57). The team imaged RCS rats at 40, 60 and 90 days postnatal, as it's known retinal degradation starts to occur at around 20 days postnatal, and found progressive degradation of the EZ line, ONL, and OPL and overall retinal thinning in RP RCS rats versus control subjects, consistent with the above results; moreover, choroidal blood flow remained constant whereas retinal blood flow was shown to decrease progressively in the RP group, becoming completely unresolvable at 60 days postnatal (57). This result is to be expected, as human RP subjects don't show choroidal blood flow changes until the end disease stages; as well, the choroid blood flow isn't metabolically regulated by the retina whereas retinal blood flow is, meaning choroidal blood flow is more likely to remain unchanged as the retinal metabolic needs shift with disease progression (58). Li's results were supported by those obtained by the second group Muir et.al, who used near identical methodology of cASL and MRI to measure blood flow and retinal thickness, but this time using a different well-established animal model; in contrast to the RCS rat, the rd10 mouse used by Muir's team had Pde6b gene mutations, which specifically hinder rod phosphodiesterase function and trigger rod cell death and cGMP build up (58). The rd10 mice also have a slightly different disease progression timeline, with retinal degradation starting at 16 days postnatal and finishing at 60 days postnatal; regardless of the slightly different disease progression, the rd10 mice showed significant retinal thinning and reduced retinal blood flow compared to control subjects, with no changes to choroidal blood flow demonstrated, mirroring the results of Li's team (58). The retinal thinning can be clearly seen in the right side of Figure 19, wherein P25 indicates 25 days postnatal, and P60 60 days postnatal given rd10 mice, relative to healthy controls on the left side. The final team led by Zhang et.al MRI imaged six human RP subjects and compared the results to those of five healthy control subjects using pseudo continuous arterial spin labelling cASL and static tissue suppression once again as was performed near-identically by Li's group, but this time in conjunction with ERG and single-shot turbo-spin-echo (TSE) acquisition to investigate blood flow; the researchers chose to image the retinal-choroidal complex as a whole versus partitioning it into retinal and choroidal blood flow as performed above, as well as aimed to deduce correlations between blood flow and ERG amplitude (35). The RP subjects imaged presented with nyctalopia, visual field constriction, blood vessel attenuation, and RPE layer changes accompanied by fundus autofluorescence, consistent with the classical RP symptoms (35). The ERG results of the RP subjects demonstrated abnormal wave profiles, undetectable profiles for rod cell responses, decreased a-wave and b-wave amplitudes, and prolonged b-waves when compared to healthy controls (Figure 18); as well, as expected, retinal-choroidal blood flow was found to be lower in RP subjects, reaching a local maximum in the retinal posterior pole (35). Zhang's team also found a statistically significant direct proportionality between ERG a-wave amplitude and blood flow; the b-wave amplitude was also positively correlated with blood flow, but not to the extent of statistical significance (35). Thus, all three times discovered reduced blood flow in the retinal-choroidal complex in RP patients, with the final team finding an additional correlation between the ERG a-wave amplitude and blood flow. It should be noted that of the above results, the data collection posture of the subject could affect blood flow, as MRI is performed with the subject lying down face up in the supine position, whereas most other optical methods involve subjects sitting up during imaging (35).

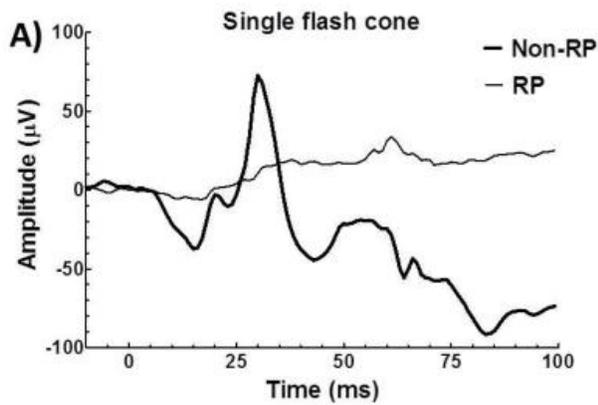

Figure 19

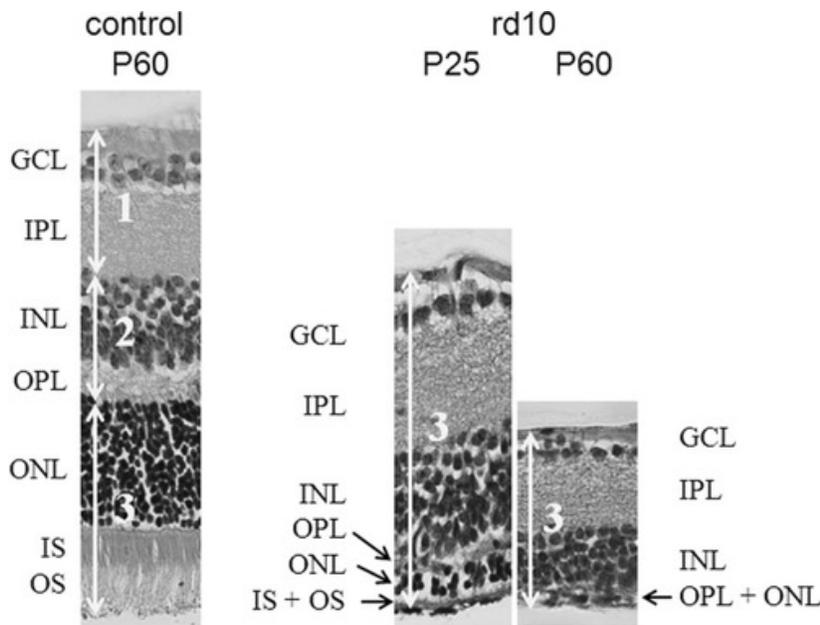

Therefore, it has been shown that each of the imaging methods of AO, OCT (including OCTA and SD-OCT), FAF, CDFI, MP, and MRI and the supplemental method of ERG provide different advantages and disadvantages when imaging RP patients, and each are best suited for measuring different parameters. Despite the diversity in results and applications of these imaging modalities, common RP diagnostic findings emerged across them all which can be used to diagnose and monitor disease progression more accurately than traditional physical examination in the absence of imaging techniques. The common features of RP as imaged by these modalities include deceased ERG amplitude, decreased EZ line width, retinal thinning, retinal blood vessel attenuation, RPE autofluorescence, the appearance and shrinking of the hyperautofluorescent ring in the fundus, and deceased retinal-choroidal blood flow. Additional physiological abnormalities such as pale optic disc pallor, cystoid macular edema (CME), and bone spicule pigmentation were also symptoms commonly detected across multiple modalities. These imaging hallmarks when combined in tandem with symptoms which present upon physical examination such as nyctalopia and narrowing field of vision can be used not just to diagnose, but also to monitor disease progression of RP. As well, it's clear different imaging methods are more suited for RP monitoring at different disease stages, with techniques such as ERG, MP, and AO better suited for early disease stages where retinal function and sensitivity degrade before physiological degradations occur, FAF and OCT better suited for mid-disease stages and onwards when morphological abnormalities appear in the retina, and CDFI and MRI better suited for end disease stages when blood flow becomes heavily impaired by vessel attenuation. With the imaging diagnosis and monitoring results from each of the aforementioned imaging techniques discussed and concluded, this review will now transition to the second to last section: the current treatments in development/in use, their application, and their respective prognoses in RP patients.

**Treatment and Prognosis of Retinitis Pigmentosa**

Though there currently exists no cure for RP, a number of innovative treatments have been developed and implemented, wherein after their application, patients show a partial but marked recovery in visual function (60). The list of treatments to be investigated is not exhaustive, and includes stem cell therapy, gene therapy, cell transplantation including RPE and omental cells, pharmacological therapy, and artificial retinal implants (60). The first treatment to be discussed, stem cell transplantation, has its efficacy rooted in the concept of supplementing pathologically damaged or lost cells with cells that have the potential to differentiate into specialized, healthy replacements (60). The direct transplantation of embryonic stem cells which have the highest differentiation potential of all stem cells in RP patients requires lifelong immunosuppressive therapy to mitigate cell rejection and teratoma formation, germ cell tumours (which are undifferentiated) and as such can contain any number of specialized tissue such as muscle and bone (60), (61). By consequence to avoid the risk of rejection, researchers often employ human induced pluripotent stem cells (hiPSCs) derived from the patient's own fibroblasts, cells which are involved in connective tissue formation and if derived from the patient themselves, won't be rejected as foreign (60), (61). These hiSCPS can differentiate into all cell types except those which are extra-embryonic, for example placental cells, and thus don't risk teratoma development nor require immunosuppressive therapy (61); hiPSC implantation and treatment success has been notably demonstrated by two groups, Li et.al (61) and Artero-Castro et.al (31), the former using mouse models, and the latter using in vitro methods, as clinical trials of stem cell therapy are still relatively novel and require further data to confer their success and safety in human subjects (61). Li's team aimed to differentiate hiPSCs into RPE cells, then transplant those cells into the RPE65 mouse, the animal model afflicted with RPE65 mutated RP, observe their result in vivo within the mice, and measure the resulting ERG response and phagocytosis; as the native RPE phagocytoses the shedding outer segments of photoreceptors (61) which occurs naturally when bright light photo-oxidizes lipids and proteins within the outer segment (29), OS phagocytosis is often a litmus test of native in vivo RPE cell function (61). The exact methodology and process through which hiSPCs are derived is a matter of biochemistry and as they aren't quite relevant to this paper, won't be discussed in detail; essentially, fibroblasts biopsied and cultured from a donor skin sample were treated with lentiviral vectors (viral vectors which can infect both dividing and non diving cells with target DNA by virtue of their nucleus-permeable pre-integration complex) to induce hiPSCs; transcription factor vectors then transduced the hiPSCs, differentiating them into RPE cells which were injected in the subretinal space of 34 newborn mice and which the mice easily integrated into the existing RPE layer architecture (61). It was found that hiPSC RPE cells demonstrated normal phagocytic function in the bovine outer segment photoreceptor samples, meaning their function was confirmed before implantation; as well, once transplanted in the mice, these hiPSCs expressed RPE markers and caused a significant recovery of ERG response and thus visual function compared to the control mice, as seen in the increased amplitude of the transplant versus control ERG waves, and the transplant wave's closer agreement with wild type (normal) ERG profiles (Figure 20) (61). Additionally advantageously, Li et.al discovered no teratoma formation within the RPE-injected mice, demonstrating the relative safety of the hiPSC method relative to that of embryonic stem cells (61). The second group who executed these pre-clinical stem cell trials in RP cases was that lead by Artero-Castro, who again investigated RPE cells derived from hiPSCs, their phagocytosis, and their potential for visual function recovery when implanted than used in vitro methods versus those of the in vivo mouse models used by Li's group; an additional contrast between the methodology of Artero-Castro's team compared to that of LI's team was their study of MERTK (Mer tyrosine kinase receptor) mutation-derived RP in contrast to the RPE65 RP mice, as well as using protein expression as a benchmark for visual function versus ERG scans (31). Their methodology overlaps with that of gene therapy, as the research team used the widely known Clustered Regularly Interspaced Short Palindromic Repeats (CRISPR) method to create and gene-correct the point mutations of the MERTK RP model, then generated hiPSCs from these gene-corrected models (31); as gene editing and gene therapy will be discussed in more detail in the proceeding subsection, it's redundant to further outline Artero-Castro et.al's gene correction process. The team then allowed the hiPSC lines to differentiate into RPE cells using a method similar to that of Li's team, introducing the proper transcription factors and nutrient medium, manipulating and allowing the RPE cells to differentiate until the correct microstructures as found in normal RPE layers were demonstrated in vitro: microvilli, cilia, and tight and adherens junctions (31). Gene expression analysis using a variant of polymerase chain reaction (PCR) proved the RPE cells expressed sufficient levels of RPE markers, as was seen in Li's results; bovine photoreceptor outer segments as were used by Li were once again employed to test and demonstrate the recovery of the RPE cell's phagocytic function (31). It was also found that the gene-corrected RPE cells derived from hiPSCs of the MERTK RP model expressed proteins in a manner consistent with the expression of the wild type (31). Thus, Artero-Castro's team adeptly showed that stem cell therapy, when paired in tandem with gene correction, can allow for the recovery of normal phagocytic function and protein expression in vitro (31). Though clinical trials of stem cell therapy in RP patients are still relatively rare compared to the pre-clinical trails above, notable studies have been undertaken by various research groups, for example that performed at the University of California Davis wherein autologous intravitreal bone-marrow stem cells are being used to treat RP in human subjects (60). Thus, stem cell therapy, as bolstered by the in vivo mouse model results and in vitro results of Li et.al and Artero-Castro et.al respectively, has been found to have a high potential for recovering the visual and normal cellular functions lost in RP disease progression. Stem cell therapy is still very much a novel treatment for genetic diseases including RP given the ethical qualms which arise from the nature of the treatment and the scarcity of clinical trial results, thus it's difficult to pinpoint the prognosis. Regardless, the pre-clinical results indicate a high degree of stem cell treatment potential for RP patients.

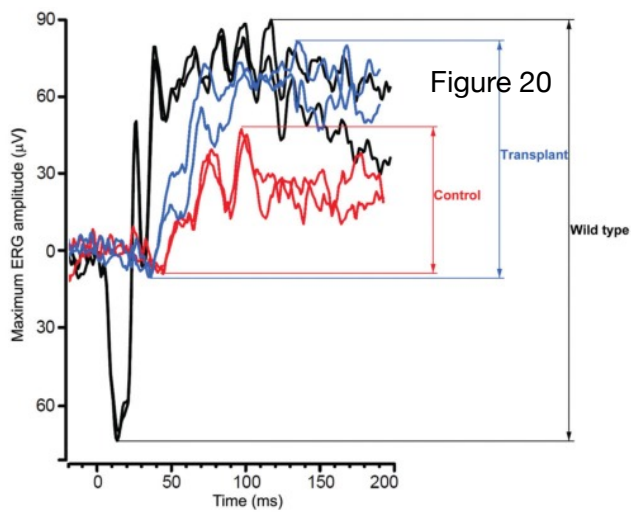

Figure 20

With the conclusion of the stem cell therapy subsection, and Artero-Castro's team using an integrated approach of both stem cell therapy and gene therapy to demonstrate the recovery of function in RP models, a natural segue into discussing gene therapy now presents. The working concept behind gene therapy involves the use of viral vectors to either correct dominant gene mutations or supplement recessive gene mutations with the normal gene; the outcome of both methods results in the genetic expression of healthy, wild type (normal) proteins (60). Many gene therapy trials use adeno-associated virus (AAV) vectors, exploiting the lack of retinal immune response and pathogenicity given AAV administration, regardless of the relatively small DNA segment of approximately 4.5 kilobases (4.5t thousand base pairs) the vector is capable of transplanting (29), (60); the most common method for treating the subject involves a subretinal injection of the target culture, in this case the viral vector housing the DNA segment of interest (60), as was seen in the subretinal injection of the hiPSC-derived RPE cells in Li's mouse models as seen above (61). Unsurprisingly, animal models were first used to test gene therapy in RP in a number of pre-clinical trials, as seen in the results of Tschernutter et.al who once again used the RCS (Royal College of Surgeon's) rat (62) and Beltran et.al who used canine models (63); though, successful clinical trials of gene-therapy treated human RP patients have also been demonstrated in recent years, as exemplified by the result of Sahel et.al (64). The first group let by Tschernutter et.al used the RCS rat (62) as was seen above in the MRI quantification of RP blood flow of Li et.al (a district group from that which conducted stem cell research on mouse models), wherein the RCS rat is the well known RP animal model with the MERTK gene autosomal recessive mutation variant (57). A lentiviral vector was again used to administer gene-correction, wherein the VSV-G pseudo types HIV-1 lentivirus containing the corrected MERTK murine cDNA segment were synthesized, then injected into the subretinal space in the same method of the above stem cell therapy results of 10 day postnatal RCS rats in only their the right eyes, maintaining the left eyes as controls (62). After monitoring the RCS rats for seven months using light stimulation and electron microscopy (microscopy which uses accelerated electrons to illuminate the sample, enabling high resolution images of tissues, cells, organelles, and even macromolecules like protein complexes), Tschernutter's group found recovery of phagocytic function, retinal function preservation, and a slowing of photoreceptor degradation (62). The fact that rod and cone cell loss wasn't reversed but rather simply slowed could be due to the fact that once the degradation process is triggered, it cannot be stopped (60).

The implication of the time-sensitive nature of gene therapy's efficacy for reversing retinal degeneration was echoed in the results of the second research group, that of Beltran et.al (63).

Beltran's team used a larger animal model, canines, afflicted with the common X-linked RPGR form of RP, a particularly the patient's fourth ORF15 gene of the the human form (the years); the canines carrying either of the two into the canine's subretinal space (63). severe variant which usually results in legal blindness by decade of life (63). Two gene point deletions in the canines have onsets which mimic the early onset of symptoms which present in the patient's first 10 were treated with an AAV 2/5 vectored human RPGR, point deletion promoters (IRBP or GRK1), then injected OCT imaging of the treated areas revealed in vivo photoreceptor preservation within the EZ and nuclei, as well as improved function relative to control eyes; as well, histopathology demonstrated correction of opsin location and photoreceptor structure, and corrected retinal remodelling at depths past the photoreceptors, with bipolar cell markers extending bipolar cell dendrites back to healthy lengths, and a thickened OPL in treated eyes consistent with normal retinal structure compared to controls (63) (Figure 21). Though, circling back to the time-sensitive nature of gene therapy, Beltran's group found subretinal AAV injection only maintained visual function and halted photoreceptor degeneration when injection occurred before appreciable photoreceptor degradations were observed, but not after degradation (63). This result implies the following prognosis: gene therapy can only be used effectively to treat RP in early retinal degenerative stages before the start of photoreceptor loss, meaning gene therapy seems to only be effective in young patients with early disease progression (63).

Figure 21

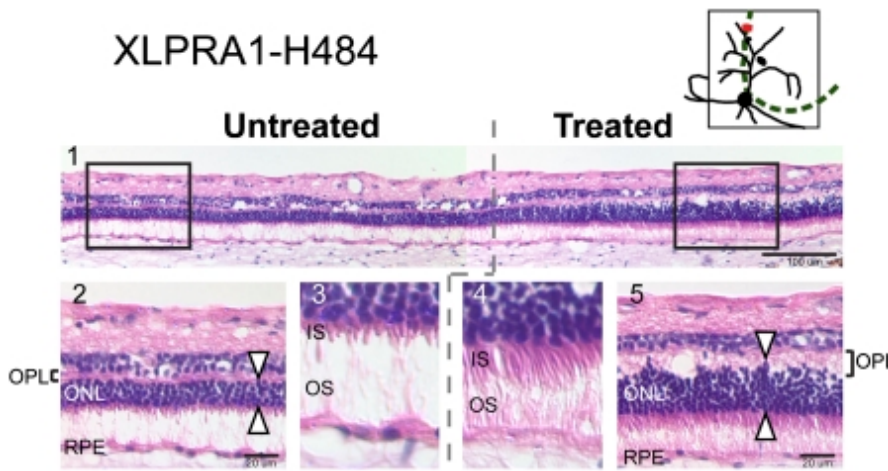

Though this time-barrier existed and was exemplified in many preclinical trials, the last group to be mentioned, Sahel et.al, found a way to overcome this limitation with their demonstration of the efficacy of gene therapy in a 58 year old male at the 40 year disease progression stage with rhodopsin-RP who was legally blind, a breakthrough which occurred quite recently in mid 2021 (64). Rhodopsin is housed within rod cells, a visual pigment through which without, photo-transduction is impossible (29). The patient when treated with Sahel's gene therapy implementation, was able to partially recover visual function, an impressive feat given his end stage of disease progression presentation (64). Sahel's group used the 2.7m8 AAV vector, which was then impregnated with ChrimsonR and fluorescent red protein, the former being a rhodopsin protein responsible for light-sensing (64). A unique innovation employed by the authors was the usage of specifically engineered goggles which the patient wore after intravitreal AAV injection into the worse visual acuity eye of the two eyes which fulfilled a triad of functions: these goggles photo-stimulated the patient's retinal ganglion cells, the target cells transduced with the AAV; captured images using integrated neuromorphic cameras, which capture movement continuously and detect intensity variations in each pixel distinctly; and finally used these intensity variations to generate images monochromatically which were then retina-projected (64). Ocular examinations were performed before and after AAV administration, including OCT and FAF for anatomical visualization; body vitals were also recorded following a general physical examination, including ECG (electrocardiogram) and EEG (electroencephalogram) tests (both analogs of ERG, wherein the former measures the electrical response from the heart, the latter the electrical response from the brain, instead of that from the retina in ERG) (64). The patient was trained visually with the goggles, and seven months post injection, started to show improvements in visual function as demonstrated by three visual and visuomotor tests; additionally, EEG tests demonstrated a visual object's presence heavily regulated the neural signals present, specifically those emanating from the occipital cortex (64). Sahel's team thus suggested that given the above results, retinal ganglion cells may be gene-corrected and photo-stimulated with the implementation of targeted genetic therapy and some apparatus capable of simultaneous light-projection and camera image capture, showing there is indeed a way to overcome the gene therapy time barrier in RP patients, given the prognosis of recovery of visual function following months after treatment in the blind patient (64). Though it should be noted, Sahel et.al gene-targeted the retinal ganglion cells (64), whereas Tschernutter et.a and Beltran et.al gene-targeted the photoreceptor cells and phagocytic function (62), (63), meaning the comparison between their approaches, results, and prognoses cannot be made in a one-to-one manner. Thus, it seems gene therapy is generally effective for treating RP, but more so in early disease stages, as its time-sensitive nature in which the prognosis heavily depends on the disease stage disallows advanced retinal degeneration from being repaired or reversed in most cases.

Following the exploration of gene therapy as a means to treat RP, the next method to be elucidated is cell transplantation, a similar yet distinct method to stem cell therapy, wherein mature cells of of interest are biopsied from tissue and transplanted, circumventing the necessity for stem cell transduction into specialized cells (60). Though, this method is often less effective due to a higher risk of adverse immune response and rejection, and less employed due to the ethical concerns of harvesting mature tissue compared to the efficacy and ethicality of stem cell therapy (60). RPE cell transplant is often the most common method in cell transplantation, as demonstrated by the work of Yamamoto et.al in an animal model (65), though the novelty of omental cell transplant in human subjects performed by Agarwal et.al as a means to treat RP (32) is also noteworthy for review. Yamamoto's group once again used the RCS rat as their RP subjects, wherein twelve rats were given electrode-guided RPE cell transplants in timeframes between sixteen and 20 days postnatal, then had their ganglion cell activity and intra-retinal ERGs (iERGs) measured and compared to the results of age-matched controls in the months following transplantation (65). Photo-stimulation of ganglion cells in the majority of the treated eyes was detected, compared to the absence of any ganglion cell activity in the controls; the treated rats also mostly demonstrated iERGs with healthy a-waves and b-waves, compared to the lack of iERGs seen in the controls, meaning RCS rats were able to preserve retinal function at the location of transplantation for timeframes of months, indicating a positive prognosis (65). Agarwal's group used cell transplants to further solidify the bright prognosis, notably in human patients, given cell transplantation, however interestingly, the research team used the fascinating implementation of omental cell transplantation for functional vision recovery in place of the commonly used RPE cells, an innovation which occurred in mid 2021 (32). The omentum is an abdominal fatty tissue structure heavily vascularized and integrated into the endocrine system which hangs from the stomach and envelops the intestines (or intra-peritoneal organs); it's known to be essential to immune system function and tissue regeneration, as the omentum has strong angiogenesis properties (the ability to form new blood vessels from existing vasculature) (32). Given omental cell angiogenesis, Agarwal et.al recognized an application of omental transplant to reverse or recover normal blood vessel morphologies in RP patients with attenuated vasculature (32). 127 RP patients underwent the procedure, in which their own omental tissue was harvested from their abdomen, then transplanted in the posterior of the eye through an abdominal tunnel leading from the gut through the chest, neck, ear, forehead, and finally eye; since the patients received transplants of their own tissue, the authors were able to avoid the possibility of tissue rejection due to normal immune response (32). The team assessed visual acuity and visual field after transplantation, as well as performed ERG, CDFI, and angiograms (cardiac blood flow X-rays). Expectedly, Agarwal's group found improved visual activity and visual field values across the transplanted patients, with the prognosis being more favourable the earlier the patient's disease progression and the higher the starting visual function parameters: CDFI showed no notable changes in extra-ocular circulation; ERG values didn't improve much, especially as the photoreceptors which had already been degraded to the point of bone-spicule pigmentation had crossed the time-sensitive threshold through which cell-based therapies are still viable in RP; and angiograms indicated no pathological retinal vasculature development (such as hemorrhages or aneurysms) and improved vasculature in the posterior hemisphere of the choroid (32). Thus, the results of both Yamamoto et.a (65) and Agarwal et.al demonstrate a marked recovery in visual function in RP patients after cell transplantation, notably either RPE cells or omental cells, bolstering the overall positive prognosis in RP cell transplant treatment despite its time sensitive nature, ethical concerns, and possibility of immune response rejection.

Concluding the investigation on cell-transplantation, the second to last treatment method to be discussed is pharmacological therapy, in other words the treatment of RP through pharmaceutical treatments (60). As previously mentioned, rod photoreceptor degeneration occurs primarily, and it's this loss which leads to cone degeneration, RPE cell death, and other retinal morphological deteriorations as a secondary effect; thus, pharmaceuticals which target rod cell preservation and survival specifically are often implemented (60); researchers primarily use AAV vectors to deliver the target compound into the retina, though encapsulated cell technology provides an alternative to viral vectors for pharmaceutical delivery (66). One of these such pharmaceuticals as implemented in RP treatment is ciliary neurotrophic factor (CNTF), a protein which promotes neural cell function, differentiation, and survival, including that of photoreceptors and ganglion cells, as demonstrated by the work of Sieving et.al (68) and Talcott et.al (67). CNTF has its receptors on the Müller glial and photoreceptor cell membranes, and has been shown to slow and prevent retinal degeneration in numerous preclinical trials involving animal models (68). The beneficial retinal effects of CNTF were taken advantage of first by Sieving's team who conducted a clinical trial of administering CNTF using surgically implanted transfected cells (enclosed cell technology) with a semipermeable cell membrane—to allow for CNTF and nutrients to outwardly diffuse, but not expose the implant to immune antagonism— in 10 RP patients over a time period of six months (68). No pathological or otherwise biochemical adverse reactions were observed in the patients, though numerous patients did present with mild fibroses on the boarders of the surgical implant, one patient developed a small posterior subcapsular cataract (PSC), and one patient developed superficial choroidal detachment (68). However, these abnormalities were shown to be either amended with the administration of corticosteroid-correction or otherwise neglected as they were shown to produce no appreciable deterioration in visual acuity (68). The results varied depending on the patient's starting visual acuity: patients with very poor starting visual acuity reported no further functional worsening, and those with better starting acuities reported improved visual acuity of greater than ten letter improvement six months post implants (68). ERG scans which were previously undetectable in patient did show to be recovered with measurable ERG flickers detected; as well, though one patient did demonstrate ERG amplitude decrease upon CNTF treatment, improvements in their visual acuity were still observed (68). ERG reduction has been shown to occur in animal models when treated with high levels of CNTF, a concurrent development to improved visual function, thus ERG scans aren't directly correlated with visual acuity in CNTF trials (68). Once the implants were safely removed, no immune cell macrophage presence was detected, meaning no adverse immune response occurred within the enclosed cells (68). Sieving's group thus found that CNTF administration resulted in general preserved and improved visual function in RP patients, though the development of abnormal retinal morphologies does dampen the prognosis of CNTF treatment, at least through enclosed cell delivery, as a viable option (68). The relative greyness of the CNTF prognosis was further supported by the work of Talcott et.al, who again used encapsulated cell technology to treat three RP patient; then, through a follow up period of several years, the team used adaptive optics scanning laser ophthalmoscopy (AOSLO), OCT, and ERG to image and measure cone density, retinal thickness, visual function, visual acuity, and visual field sensitivity (67). Cone photoreceptor tracking with AOSLO revealed no appreciable deteriorations in cone density occurred after CNTF treatment, as seen in the consistent spacing between the cones (red plus signs) in Figure 22 over the data collection period, implying CNTF promotes cone survival (67). However, no noticeable improvements were seen in visual acuity, visual field sensitivity, and ERG scans (67). Thus the work of Sieving and Talcott both demonstrate though CNTF is known to slow and present photoreceptor loss, its prognosis isn't as positive as those of the treatments about as its usage doesn't always result in improved visual function, as only some patients of Sieving's trial reported visual acuity improvement (68), and Taclott's patients showed no visual improvement despite cone density preservation (67).

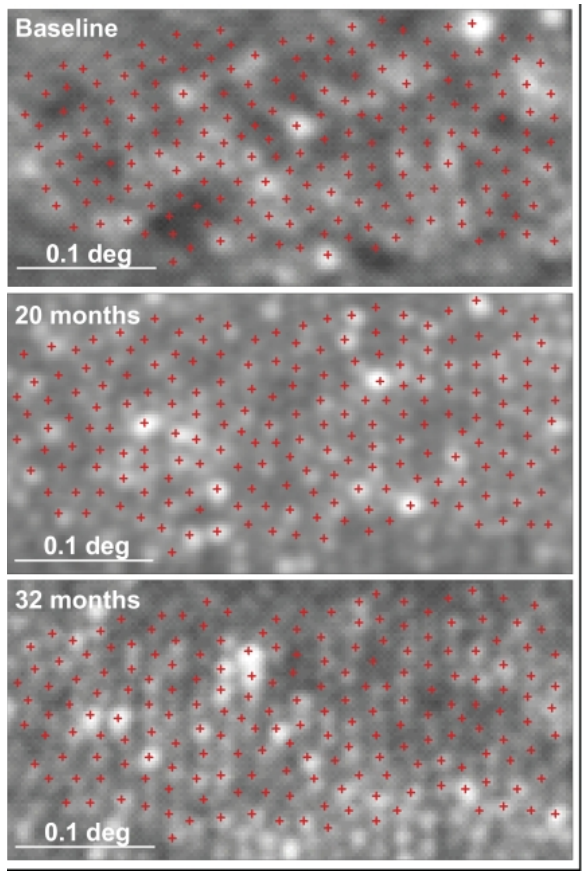

Further pharmaceutical trials have also been performed using different factors, for examples those involving rod-derived cone viability factor (RdCVF), as seen by the preclinical trial results of Byrne et.al (69). As it was mentioned briefly above, the physiological mechanism by which primary rod loss triggers secondary cone loss is still slightly ambiguous, though one such inference of this degradation involves the usage of RdCVF: the factor is normally secreted by healthy rod cells, but becomes scarcer as RP progresses and ravages the rods, thereby dissuading cone survival (69). Byrne's group used AAV vectors to implant two RdCVF isomers, RdCVF and RdCVFL, into the rd10 (retinal degeneration 10) mouse, a widely used animal model for autosomal recessive RP; the AAV-RdCVF complexes were injected either through the tail vein on the first day postnatal or through the intravitreal area at fifteen days postnatal (69). Some of the mice were dark-reared, meaning they developed in visually deprived or dark conditions to impair visual cortex development to observe if cone rescue given RdCVF treatment could occur (69). It was found in general that ERG amplitudes did increase slightly in treated mice versus control subjects; as well, concurrent expression of both RdCVF isomers was shown to rescue cone cells more effectively than single expression of RdCVFL (69). Interestingly once again, ERG was shown not to be a direct measure of visual function recovery in the context of pharmacological treatment (69). Additionally, the dark-reared mice demonstrated decreased oxidative stress, increased rhodopsin, and extended rod survival when RdCVFL was expressed early on in their life cycles (69). Thus, Byrne et.al showed that both RdCVL isomers preserve photoreceptor function separately, yet still in a manner in which some form of co-expression or cooperation occurs between the two (69). The three authors were able to demonstrate though there exists a high degree of potential and efficacy in pharmacological RP treatment, the prognosis and viability of the treatment is still significantly less beneficial than that of other treatments like gene therapy and stem cell therapy; pharmacological treatment has shown less conclusive results, and poses the risks of pathological immune response or abnormal morphology formation depending on how the treatment was administered.

In some patients with end-stage RP, the above treatments simply aren't efficacious as the pathological progression is too advanced; in these cases, the final treatment method can be implemented; electronic retinal implantshave been shown to recover vision in circumstances which seem dire (60). Two such groups which implemented retinal implants were Dorn et.al (70) and Castaldi et.al (72). The first group led by Dorn used a specific retinal prosthesis, 60-electrode Argus II, and implanted them in 28 fully blind RP patients (defined as worse than or equal to bare light perception) in the end stage of disease to see if the patients could detect the directional motion of a bar scrolling across a screen in high contrast (70). The central tenet of retinal electronic implant efficacy is the implantation of electrode arrays which can explicitly, electrically stimulate surviving retinal cells in the inner retina, specially the bipolar and ganglion cells, to recover some form of visual function (70). The Argus II retinal prosthesis used by Dorn's group is an array of 60 electrodes which function independently, along with a camera, video processing hardware, and an inductive coil for wireless data and power transmission, which was implanted in the epiretinal region, the fibrous or scar tissue on the macula, of the patients (70). Other popular methods of implantation include subretinal and suprachoroidal pathways (71). The prostheses detect video, digitizes the video, then condenses the data into a 60 pixel grid in which each pixel corresponded to one electrode's stimulation, from which optical sensitivity can be deduced (70). The results were encouraging, but not conclusive: of the 28 patients, fifteen were able to detect the direction of the bar's motion much more accurately with the prosthesis on than with it off; eleven were incapable of identifying directional motion, with the prosthesis on or off; and two seemed to perform more accurately when the system was off (70). Though regardless of the relative inconsistency of the results, Dorn's work is impressive, as it demonstrates clinical evidence of visual recovery in end-stage human RP patients, a feat which is far from easy to accomplish; as well, no pathogenic or negative immune responses occurred after implantation, nor did any abnormal morphologies form, indicating the positivity of the prognosis of RP patients with highly advanced disease progressions, improvements which couldn't otherwise occur with the administration of any of the above treatments given the harsh and irreversible advancement of retinal degeneration. The second group, Castadi et.al, further exemplified the remarkable effects artificial retinal implants can have on late stage RP patients' vision (72). Castaldi's group once again used the Argus II retinal prosthesis to test for visual function (defined again in terms of bare light

Figure 23

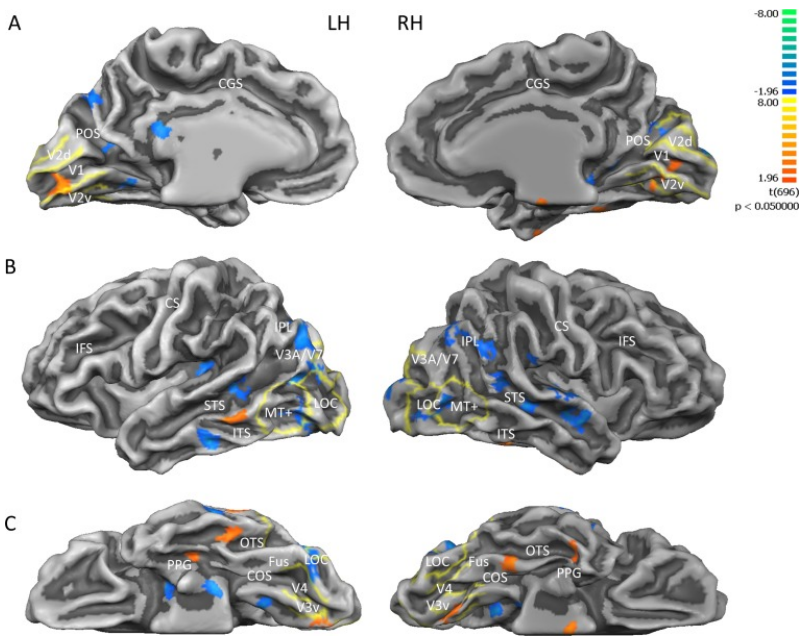

perception) RP patients, using functional MRI to measure blood oxygen saturation as a means to infer neural activity and overall neuroplasticity (72). As both authors used Argus II, it should be noted the exact mechanism by which this prosthesis, and indeed prostheses of a similar nature, actually restore vision isn't completely understood; ambiguity exists on whether or not the restoration is due to retinal activity or neural activity in the brain (72). Regardless of this slight opacity in the functional mechanism, Argus II retinal implantation has nonetheless shown to be effective in recovering some degree of vision, as was demonstrated by the ameliorated Blood

Oxygenation Level Dependent (BOLD) visual responses in the thalamus and cortical of the brain in patients after implantation (72). The prostheses were implanted in the choroid, with no physiological complications seen except for in one patient who developed a mild choroidal detachment which was quickly corrected with surgical intervention (72). Over a period of two years following implantation, the patients were trained to use the Argus II device, with subtle but significant improvements in BOLD responses seen across all patients in the V1, V2, and V3 brain cortices, especially in those who trained more intensely with the device, as seen in Figure 23 in which the blue regions represent neural response before implantation, and the orange and yellow regions represent neural response after implantation (72). The BOLD response improvements bolster not only the efficacy of Argus II retinal implants as a means for treating RP, but also the remarkable neuroplasticity within the adult brain, even after years of stimuli absence as seen in these adult patients with highly advanced RP disease stages (72). It should be noted numerous other studies and clinical trials have reported on the fairly consistent efficacy of electronic retinal implants, however those results involved treatment of other blindness-causing pathologies like Stargadt's disease (in which fatty deposits build up on the macula) and age-related macular degenerations (AMD, a self-explanatory pathology); as such, though they cement the benefits patients can reap after receiving retinal electronic implants, these results won't be shared as they aren't relevant to RP treatment. Thus in general, it seems retinal electronic implants are a useful means of recovering visual function, providing a highly positive prognosis and hope for RP patients with such severely deteriorated retinas that other treatments aren't viable because the retinal infrastructures which these treatments operate upon simply aren't there anymore. As well, though there exists the potential for morphological complications, for example the choroidal detachment seen by Castaldi's group, these complications can be quickly resolved with overall benign effects on the patient (72). Therefore, the most popular, current treatment methods for treating RP including stem cell therapy, gene therapy, cell implantation, pharmacological treatment, and retinal electronic implant, have been discussed, including their general implementation and prognoses. Though there currently exists no widely agreed upon RP treatment, further work on these innovative methods provides much potential promise for someday creating a cure.

**Conclusion**

In conclusion, this paper has reviewed retinitis pigmentosa in terms of its genetic causes, the medical imaging modalities most commonly used to study it, the symptoms and diagnostic imaging results, and the various treatments currently in development to combat it. The various types of pathological mutations were discussed, including autosomal dominant, autosomal recessive, and X-linked RP, and retinal physiology was traced for context; following the brief interlude on genetics and disease cascades for understanding RP's cause came a background of retinal physiology and medical imaging theory and methodology, including those on adaptive optics, various OCT implementations of spectral-domain OCT and OCT angiography, fundus autofluorescence and fluorescence lifetime imaging ophthalmoscopy, colour Doppler flow imaging, microperimetry, and MRI with a succinct section on ERG. Next, the symptoms and diagnostic imaging results of each imaging modality were traced, with emphasis on the idea that macroscopic visual defects a patient may experience upon physical examination require genetic testing and more usually medical imaging as a means to confirm diagnosis. The hallmark symptoms included narrowing field of vision, nyctalopia, waxy optic disc colour, bone spicule pigmentation, the appearance and shrinking of hyperfluorescent rings in the fundus, shortened ellipsoid zones of the photoreceptors, retinal thinning, and blood vessel attenuation. It was also found that different imaging techniques are more suitable for visualizing and monitoring different disease stages: adaptive optics, microperimetry, and ERG were well-tuned for early disease stage monitoring when morphological abnormalities have yet to be detected, yet visual and retinal function and sensitivity have started to deteriorate; fundus autofluorescence and OCT were well-tuned for mid-disease stages when retinal degeneration is advanced enough to result in fluorescent fundus deposits and other physiological deformities such as a reduced ellipsoid zone line; and CDFI and MRI were well-tuned for end stages of the disease in which vessels start to attenuate and impair blood flow. Then, the various treatments in development and their respective prognoses were investigated, including stem cell therapy, gene therapy, cell transplantation, pharmacological therapy, and electronic retinal implants. Through the discussion of the treatments, it was observed that the efficacy of any RP treatment is highly time-sensitive, wherein once retinal degeneration has reached a certain point, it cannot be reversed. Stem cell and cell transplantation therapies both pose ethical issues but have a high degree of efficacy; though, the prognosis of the latter is less completely beneficial as it often results in the formation of harmful retinal physiologies around the transplant site. Gene therapy was found to provide a positive prognosis in patients when the pathological mutation is corrected or otherwise supplemented, whereas the efficacy of pharmacological therapy isn't fully conclusive. Finally, electronic retinal implants were shown to help patients in highly advanced disease stages recover some form of visual function. Thus, with the conclusion of this paper, retinitis pigmentosa has been elucidated in its causes, how we image it, and how we treat it. The essentiality of continuing to develop and advance medical imaging techniques and treatments is clear. And though treatments have yet to reach a consensus on what's accepted, the continual refinement and research of treatment methods in conjunction with the simultaneous advancement of medical imaging techniques provide hope of early intervention and the development of a cure for this genetic blindness that affects so many around the world.